\documentclass[times,3p,review,12pt]{elsarticle}
\usepackage[utf8]{inputenc}
\usepackage[english]{babel}
\usepackage{array}
\usepackage{gensymb}
\usepackage{caption}
\usepackage{float}
\usepackage{mathtools}
\usepackage{natbib}
\usepackage[abs]{overpic}
\usepackage{subfigure} 
\usepackage{booktabs,caption}
\usepackage[flushleft]{threeparttable}
\usepackage{setspace}
\usepackage[inline]{enumitem}
\usepackage{tabu}
\usepackage{multirow}
\usepackage{enumitem}
\usepackage{scrextend}

\usepackage{booktabs, siunitx}
\usepackage{url}
\usepackage[tableposition=below]{caption}
\usepackage{gensymb}
\usepackage{graphicx}
\usepackage{amsmath}
\usepackage{amsmath,bm}
\usepackage{amssymb}
\usepackage{commath}
\usepackage{setspace}
\usepackage{geometry}
\usepackage{textcomp} 
\numberwithin{equation}{section}
\usepackage[toc,title,page]{appendix}
\usepackage{xcolor}
\usepackage{titlesec}

\titleformat{\paragraph}
{\normalfont\normalsize}{\theparagraph}{1em}{}
\titlespacing*{\paragraph}
{0pt}{3.25ex plus 1ex minus .2ex}{1.5ex plus .2ex}

\usepackage{hyperref}
\hypersetup{
    colorlinks =    true,
    linkcolor =     [rgb]{0.9,0.4,0.4},
    anchorcolor =   [rgb]{0.9,0.4,0.4},
    citecolor =     [rgb]{0.4,0.4,0.4},
    filecolor =     [rgb]{0.4,0.4,0.4},
    urlcolor =      [rgb]{0.4,0.4,0.4},
    pdftitle=       {Title},
    pdfsubject=     {Title},
    pdfauthor=      {A. Uthor}
}
\usepackage[nameinlink]{cleveref}
\crefname{figure}{Fig.}{Figs.}
\crefname{equation}{Eq.}{Eqs.}
\crefname{section}{Section}{Sections}
\crefname{table}{Table}{Tables}
\Crefname{figure}{Figure}{Figures}
\Crefname{table}{Table}{Tables}
\Crefname{section}{Section}{Sections}
\Crefname{equation}{Equation}{Equations}

\makeatletter

\newcommand{\vast}{\bBigg@{4}}
\newcommand{\Vast}{\bBigg@{5}}

\makeatother

\setcitestyle{authoryear,open={(},close={)}}
\usepackage{framed} 
\usepackage{multicol} 

\usepackage{nomencl} 
\usepackage{xpatch}

\newlength{\nomitemorigsep}
\makenomenclature

\renewcommand*\nompreamble{\begin{multicols}{2}}
\renewcommand*\nompostamble{\end{multicols}}
\usepackage{ifthen}
\renewcommand{\nomgroup}[1]{%
 \itemsep\nomitemorigsep%
  \ifthenelse{\equal{#1}{A}}{\item[\textit{Bulk material properties}]}{%
  \ifthenelse{\equal{#1}{B}}{\item[\textit{Bed parameters}]}{%
  \ifthenelse{\equal{#1}{E}}{\item[\textit{Subscripts}]}{%
  \ifthenelse{\equal{#1}{D}}{\item[\textit{Other symbols}]}{%
  \ifthenelse{\equal{#1}{F}}{\item[\textit{Abbreviations}]}{%
  \ifthenelse{\equal{#1}{C}}{\item[\textit{Interparticle parameters between particles $i,j$}]}{}}}}}}%
 \itemsep\nomitemsep
}

\begin{document}
\begin{frontmatter}

\title{
Optimized mouldboard design for efficient soil inversion using the discrete element method
}
\cortext[equal]{Equal contribution}
\cortext[cor]{Corresponding author: ratna@iitm.ac.in, Tel: +91-44-2257-4719, Fax: +91-44-2257-4652}
\author[add1]{Vinay Badewale\corref{equal}}
\author[add1]{Sujith Reddy Jaggannagari\corref{equal}}
\author[add2]{Prasad Avilala}
\author[add1]{Ratna Kumar Annabattula\corref{cor}}
\ead{ratna@iitm.ac.in}

\address[add1]{Mechanics of Materials Laboratory, Department of Mechanical Engineering, Indian Institute of Technology Madras, Chennai- 600036, India}
\address[add2]{Altair Engineering India Private Limited, Bengaluru-560001, India}

\begin{abstract}
The design of a mouldboard (MB) plough is critical for achieving efficient soil inversion, which directly impacts soil aeration, weed control, and overall agricultural productivity. In this work, a design modification of the cylindroid-shaped MB plough is proposed, focusing on optimizing its surface profile to enhance performance. The discrete element method is used to simulate the ploughing process and evaluate the performance of the modified plough profile. The modified plough profile is compared against a previously proposed design to assess its impact on soil inversion efficiency, wear reduction, and stress distribution. A novel methodology is introduced to evaluate the plough’s performance in soil inversion. The modified design demonstrates superior soil inversion efficiency, with improvements of up to 32.95\% in the inversion index for different velocities. The modified design achieves a notable reduction in wear up to 23.7\%, compared to the original design. Although a slight increase in stress is observed in the modified design due to higher forces, the induced stresses remain well within the permissible limits for the plough material. Overall, the findings highlight the advantages of the modified plough design, including enhanced soil inversion efficiency and reduced wear, underscoring its potential for improved performance in tillage applications. However, the current study is limited to simulation-based analysis without experimental or field validation. Future work will focus on full-scale physical experiments to validate the simulation outcomes and incorporate additional factors such as depth-dependent moisture, soil cohesion, and multi-factor wear models for improved predictive accuracy.

\end{abstract}
\begin{keyword}
Mouldboard plough\sep Soil inversion\sep Discrete element method \sep Finite element analysis \sep Soil-tool interactions
\end{keyword}
\end{frontmatter}
\nomenclature{$H_{T}$}{Mass fraction of top layer in region 2}%
\nomenclature{$H_{B}$}{Mass fraction of bottom layer in region 3}%
\nomenclature{$V_{T}$}{Mass fraction of top layer in region A}%
\nomenclature{$V_{B}$}{Mass fraction of bottom layer in region C}%
\nomenclature{$\alpha$}{Cutting angle that the cutting edge (AB) in plane YOZ make parallel to the travel direction OY}%
\nomenclature{$\beta$}{Lateral vertical angle that the cutting edge makes with the horizontal in a vertical plane (XOZ) perpendicular to the travel direction}%
\nomenclature{$\gamma$}{rake angle between the wedge face (ABC) and the horizontal plane including the cutting edge}%
\nomenclature{$\theta$}{lateral directional angle that the cutting edge makes with the travel direction in a horizontal plane}%
\nomenclature{$\Delta \theta$}{change in angle as the generatrix moves from the bottom to the top of the surface}%
\nomenclature{$p$}{parameter of the parabola}%
\nomenclature{$q$}{vertical translation of the parabola vertex}%
\nomenclature{$a$}{furrow height}%
\nomenclature{$b$}{furrow width}%
\nomenclature{$H$}{maximum height of the tool}%
\section{Introduction}\label{Sec:Intrduction}
Agriculture is an essential part of the global economy and is crucial for ensuring global food security. However, the sector faces significant challenges as urbanisation accelerates, indicating a decreasing interest in farming among the population. This demographic shift necessitates modernising and enhancing agricultural practices to meet the increasing demands of urban populations. Ploughing is the initial procedure in a farming field and plays a vital role in determining crop production quality. Tillage is the mechanical manipulation of soil and plant residues to prepare a seedbed for crop planting \citep{Reicosky2003}. Tillage is categorised into primary and secondary types. Primary tillage breaks up compacted soil, buries debris and weeds, and pulverises the soil to facilitate sowing. Secondary tillage achieves the desired soil texture for optimal planting conditions.

The mouldboard (MB) plough is a standard primary tillage equipment that performs four functions: cutting, elevating, inverting, and pulverising the soil furrow slice. It is one of the most costly and energy-intensive aspects of agricultural production, accounting for 25-44\% of global greenhouse gas emissions. A study on fuel consumption and the time needed to prepare the seedbed for various tillage implements found that preparing a seedbed requires the most time and energy when using a mouldboard plough \citep{Pratibha2019}. Simulation-based design can help speed up the design and performance evaluation of mouldboard tillage equipment. Designing an optimised mouldboard plough has been a study of interest for many years. As a result, several researchers have devised various mouldboard designs, such as cylindrical, cylindroid, helical, and the most popular cylindroid shape. 


The ploughing process can be investigated in three ways: empirical, analytical, and numerical. Several researchers have modelled the soil as a continuum and simulated the ploughing process by finite element method \citep{Bentaher2013,Asaf2007,Ibrahmi2013}. The interaction between soil particles and between soil particles and rigid or flexible bodies can be modelled using the discrete element model (DEM) \citep{Cundall}. DEM is used to simulate the behaviour of granular materials and may be used to optimise the design process. The separation and mixing of soil layers, the development of cracks, and the flow of soil particles during soil deformation, particularly during the tillage process, cannot be adequately modelled by the finite element method \citep{C.Plouffe1999}. Determining soil properties and soil-tool interaction parameters remains a critical challenge in this numerical method. The DEM approach proves to be highly effective in addressing this issue \citep{zeng2025parameter,zeng2024numerical,lan2024simulation}.

According to \citet{RosVSmithRJMarleySJ1995}, the development of suitable ways to define the geometry of tillage tools remains a fundamental challenge in tillage tool design. \citet{V.Craciun1998} proposed a mathematical model employing parametric equations which can be used to determine the surface of the cylindroid mouldboard plough, which can predict the forces on the ploughing surface. Due to the large number of parameters and shape design using parametric equations, optimisation of such a model becomes complicated and time-consuming. \citet{Godwin2007} developed a mathematical model that can predict the draft force and its components acting on the parts of the mouldboard, followed by experimental validation. \citet{Formato2017} performed the design optimisation of the MB plough surface by the computerised mathematical model. However, analysis of the soil inversion after the tillage has not been studied yet. Soil inversion plays a critical role in enhancing agricultural productivity by influencing several key soil and crop parameters. Effective inversion leads to improved soil aeration, which enhances oxygen availability for root respiration and microbial activity. It also facilitates better root penetration by loosening compacted soil layers, thus improving water and nutrient uptake. Additionally, soil inversion aids in the incorporation of crop residues and organic matter, which contributes to soil fertility and structure. Importantly, the process also helps in weed control by burying weed seeds and disrupting their germination cycle. These combined effects promote healthier crop growth and higher yields. Therefore, optimizing the soil inversion process through improved tillage tool design is essential, and numerical simulation offers a powerful approach to evaluate and enhance these design parameters under controlled and reproducible conditions.

Several researchers have investigated the forces acting on the mouldboard plough during soil-tool interaction. In previous studies, vertical, draft, and lateral forces were determined using mathematical models. The variation of reaction forces has been studied with varying ploughing velocity by \citet{Suministrado1990}. 
Based on parametric equations given by \citet{Ibrahmi2017b}, a comparison study of cylindroid and cylindrical mouldboard ploughs was conducted. Later this work was extended to study the draft and energy requirement to drive the plough using FEM \citep{Ibrahmi2017,Bentaher2013,Azimi-Nejadian2019}. A comparison of a proposed new graphical approach of mouldboard plough bottom with earlier graphical methods was undertaken. The weight of the mouldboard bottom was reduced by 7.3\% using the new approach by \citet{ShahmirzaeJeshvaghani2013a}. \citet{Mun2011} worked on describing and characterising the 3D plough based on the spline concept to improve the classic MB plough that had been built previously. \citet{McKyes1997} studied the effect of design parameters of the tillage tool, which is a flat rectangular shape, performed on the loosening of moist clay soil.

\citet{Ucgul2016} examined the influence of the coefficient of rolling friction and cohesion energy density on the angle of repose and soil-tool force. \citet{Ucgul2017} used DEM to model the mouldboard ploughing and studied the topsoil burial at a full scale under field conditions. Researchers have tried to examine and modify the shape using the bionic design \citep{Li2016,Sun2018}. The wear of the tool is one of the significant parameters that affect the tool's life. Various models have been proposed to predict the wear of the surface of the tool, and the most widely used are Finnie wear \citep{Finnie1960} and Archard wear \citep{Archard1953}. However, DEM only calculates the virtual wear numerically, and the material removal from the tool surface is not considered. \citet{Owsiak1997} investigated and developed a mathematical model to estimate the wear on the symmetrical wedge-shaped tillage tool made up of steel experimentally. \citet{Hoormazdi2019} predicted the soil-tool abrasive wear by combining numerical and experimental approaches. \citet{Schramm2020} modelled wear via scratch test using DEM, which studies the influence of the continuous wear on the tool's surface. \cite{zhang2024simulation} integrated finite element simulations with field experiments using the Archard wear model, offering a predictive framework for assessing wear in rotary plough blades under realistic operating conditions. \cite{yao2025wear} provided a theoretical foundation for optimizing the structural design and material selection of plough tips.

While several studies have investigated the mechanical performance of tillage tools through field testing and empirical modeling, limited attention has been paid to the numerical simulation of wear mechanisms under realistic soil conditions. From the literature, most existing work either focuses on soil inversion performance or estimates wear using simplified assumptions, without coupling the two phenomena in a single simulation framework. Recent developments in Discrete Element Method (DEM) have enabled more detailed modeling of soil-tool interactions; however, the incorporation of dynamic wear modelling, particularly in the context of moldboard ploughs, remains underexplored. Furthermore, comparative assessments of novel plough designs with conventional tools using simulation-based metrics such as stress distribution, draft force, and wear index are scarce. This study aims to address these gaps by employing a DEM-based approach to evaluate a previously developed moldboard plough design, with an emphasis on wear behavior, soil inversion efficiency, and stress distribution under controlled conditions. The profile of the mouldboard plough presented by \citet{Ibrahmi2017b} is modified by a graphical method and shifting the turning point to the edge. A new design parameter to investigate the inversion capacity of the tool is proposed, and it is called the inversion index. The inversion index, wear, stress and forces are compared for both designs.

\section{Methodology} \label{Sec:Methodology} 
\subsection{Designing of mouldboard plough}\label{Sec:Tool_design}
In this work, the mouldboard plough design of \citet{Ibrahmi2017b} is the baseline design.  The design of the MB plough can be divided into three steps, as illustrated in \Cref{Fig:Exp_Setup1}.
\begin{enumerate}
    \item Basic geometrical parameters of the elementary triangular wedge
    \item Surface generation. 
    \item Profile generation. 
\end{enumerate}
\subsubsection{Basic geometrical parameters of the elementary triangular wedge}
An elementary triangular wedge OABC with spatial plane surface ABC can describe a tillage tool, as shown in \Cref{Fig:Exp_Setup2}. The direction of the tool travel is along the OY axis. The triangular wedge consists of mainly four angles, i.e., cutting angle ($\alpha_{0}$), lateral vertical angle ($\beta_{0}$), rake angle ($\gamma_{0}$), and lateral directional angle ($\theta_{0}$). The soil layer is cut by a triangular wedge, as shown in \Cref{Fig:Exp_Setup2}. The cutting edge, denoted as AC, has its cutting point A as the initial point of contact with the soil slice.

The separation of the soil can be understood by taking components of the soil in three different directions. The components of soil flow over the tool surface are represented by flow 1, flow 2, and flow 3. Flow 1 is along the edge AB, which means the lifting capacity of the wedge and is determined by $\alpha_{0}$. The lateral direction angle $\theta_{0}$ separates the soil layers on one side and is denoted by flow 2. Similarly, vertical lateral angle $\beta_{0}$ affects the flow of the soil particles vertically upwards perpendicular to the tool travel direction and is denoted by flow 3. The combination of flows 1, 2, and 3 represents the movement of the entire soil layer over the face ABC. The relationship among the four angles that define the triangular wedge is given by \cite{RosVSmithRJMarleySJ1995},
\begin{equation} \label{eq: relation between four angles}
\tan \alpha_{0}=\tan \gamma_{0} \sin \theta_{o}=\tan \beta_{0} \tan \theta_{o}
\end{equation}
\subsubsection{Surface generation}
\begin{figure}[H]
\centering
\subfigure[]{
\includegraphics[clip, trim = 0.0 -1.5cm 0. 0.0, width=0.5\textwidth]{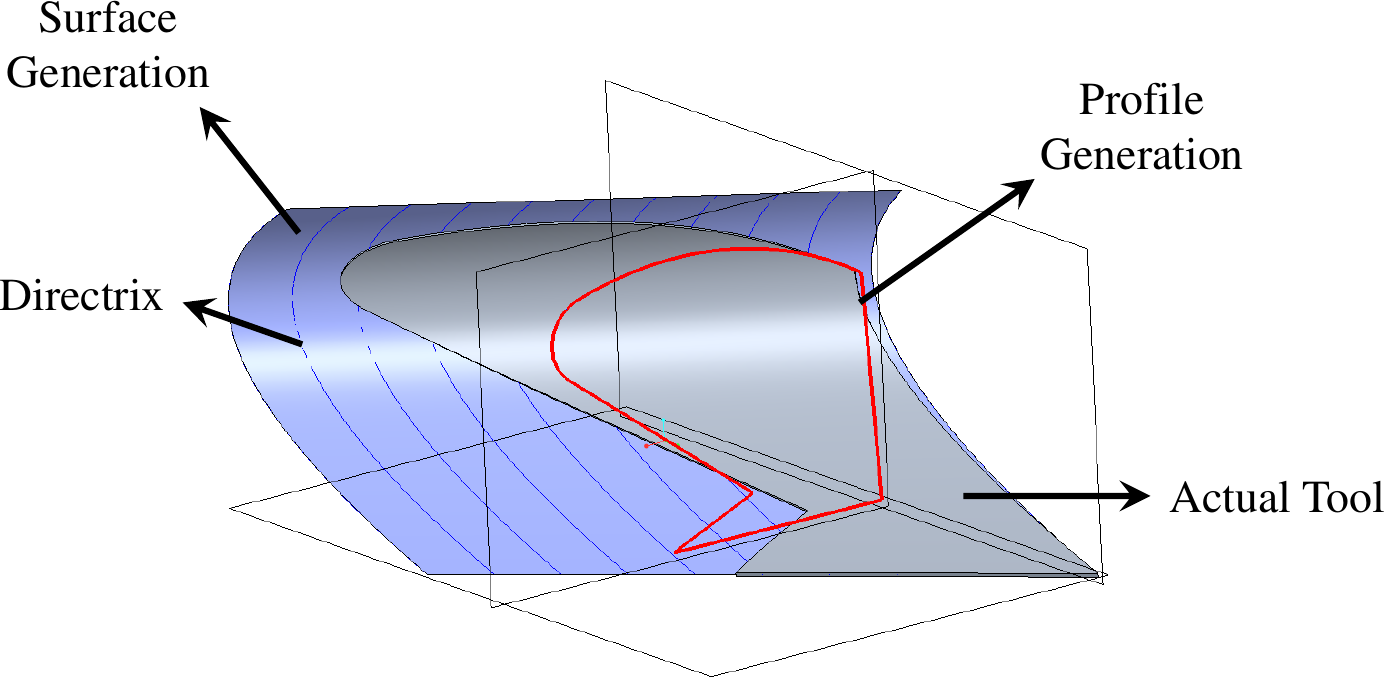}
\label{Fig:Exp_Setup1}}%
\subfigure[]{
\includegraphics[width=0.5\textwidth]{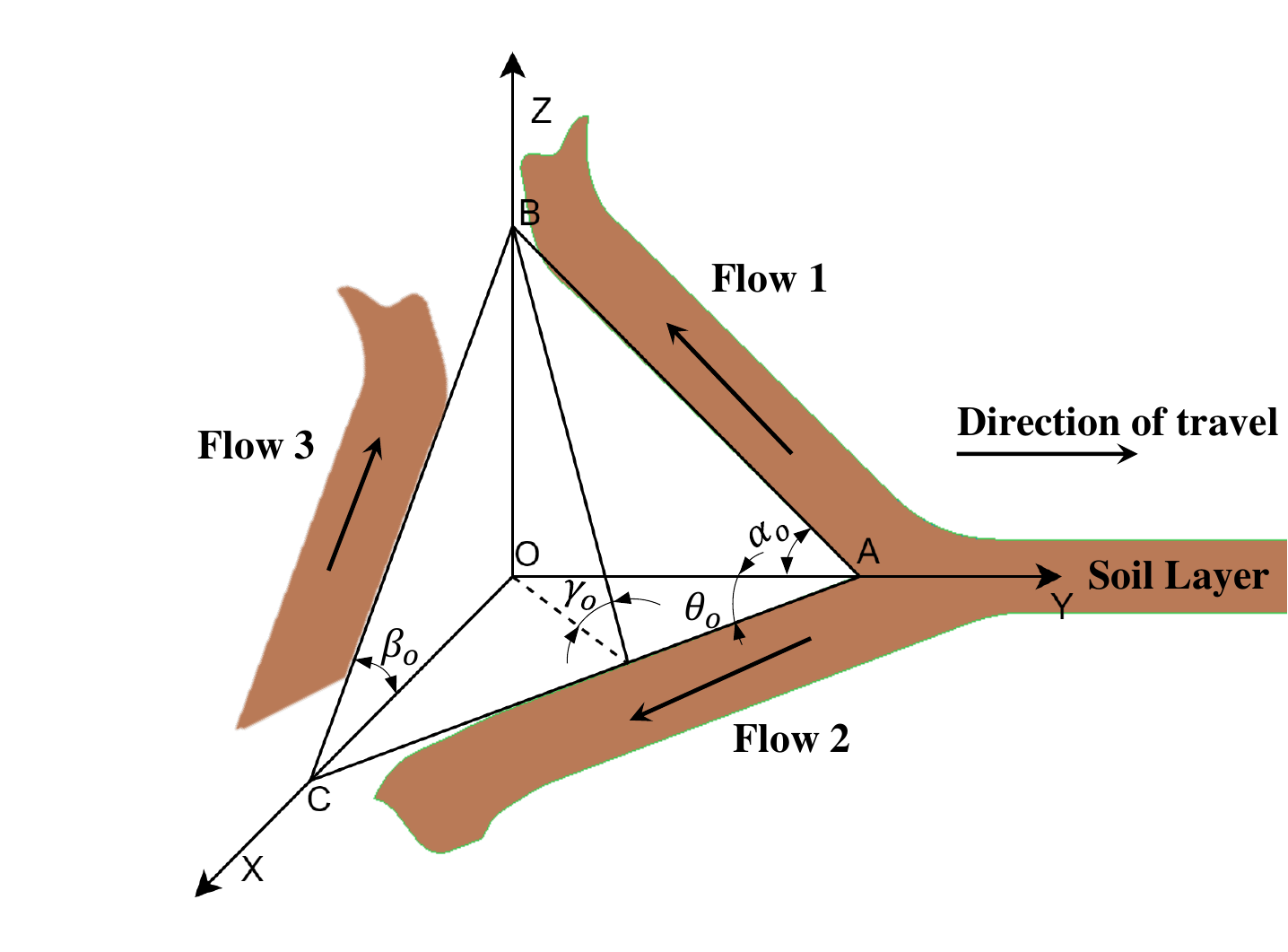}
\label{Fig:Exp_Setup2}}
\subfigure[]{
\includegraphics[width=0.6\textwidth]{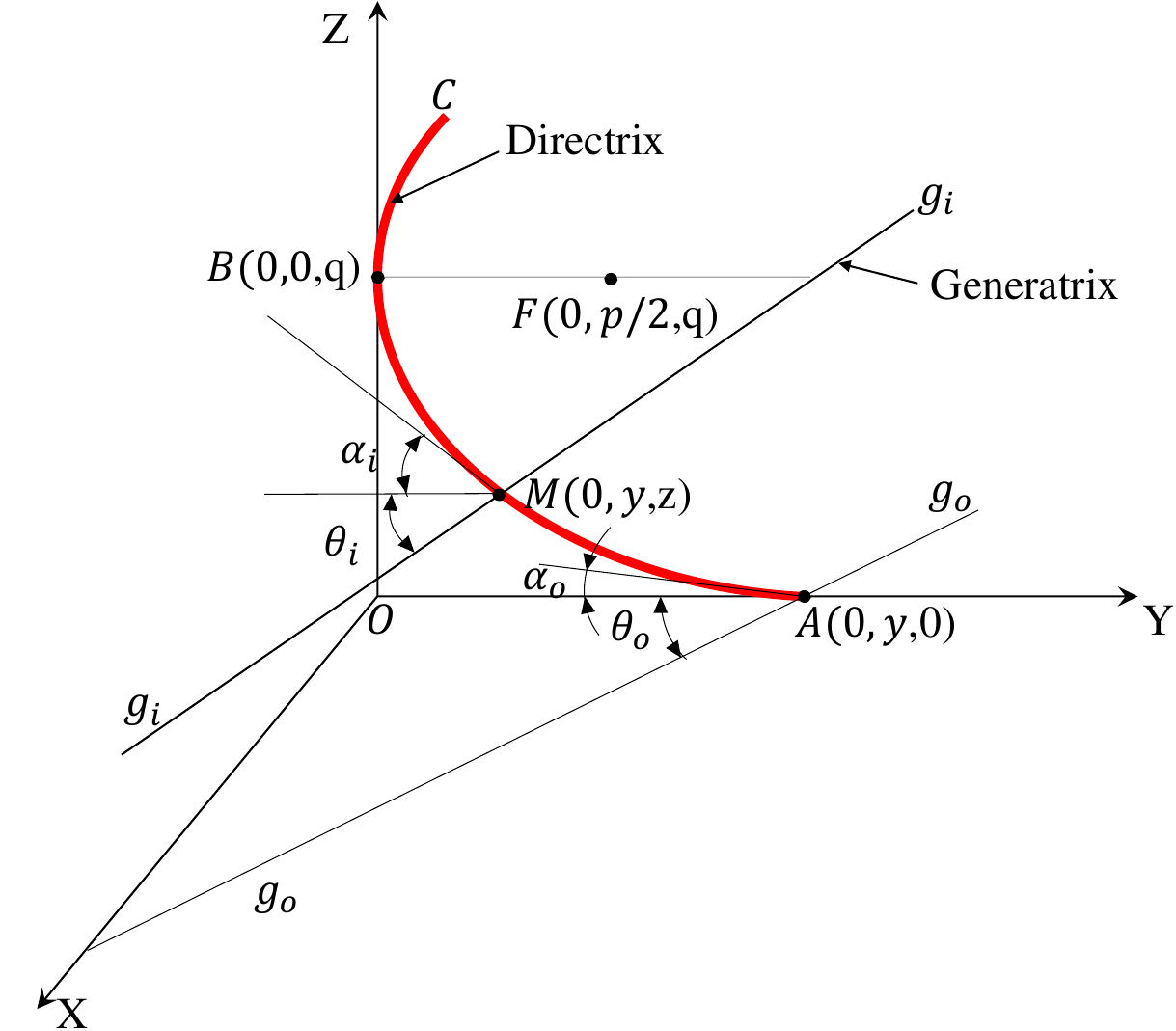}
\label{Fig:Exp_Setup3}}%
\caption{(a) Procedure to design MB plough, and (b) triangular wedge for defining the geometrical parameters of a passive tillage tool, (c) tool surface generation}
\label{Fig:1}
\end{figure}

When a curve, referred to as the directrix, moves along a specified path, known as the generatrix, it generates a surface. The method for constructing the directrix, a parabola, is described here. The parabolic curve (AC) is a function of three independent angles, i.e., $\alpha_{0}$, $\gamma_{0}$, $\theta_{0}$ and geometric parameters $p$ and $q$ as shown in \Cref{Fig:Exp_Setup3}. The $p$ and $q$ are geometrically related as,
\begin{equation} \label{pq relation}
p=q \tan \alpha_{0}
\end{equation}

The generatrix is represented by $g_{i}$ at a \textit{$i^{th}$} position on the OZ axis. The intersection of the generatrix and the directrix is denoted by a generatrix ($g_{i}g_{i}$) is always parallel to the horizontal plane XOY. The equation of the directrix is given by \cite{RosVSmithRJMarleySJ1995},
\begin{equation} \label{lit equation}
y_{i}=\dfrac{(z-q)^{2}}{2 p}-\dfrac{x_{i}}{\tan \left(\theta_{0}+\dfrac{z^{2}}{2r}m\right)}
\end{equation}
The term $\left(\theta_{0}+\dfrac{z^{2}}{2r}m\right)$ represents the variation of $\theta_{0}$ along the height of the tool. The parameter $m$ is the scale coefficient expressing the increment in the $\theta$ with respect to $z$. Where $\theta$ is the important angle of the mouldboard surface, as shown in \Cref{Fig:Exp_Setup3} and $z_{i}$ is the height of the point $i$ and given by,
\begin{equation}
m=\dfrac{\theta_{\mathrm{max}}-\theta_{\mathrm{o}}}{\mathrm{z}_\mathrm{max }^{\prime}}=\dfrac{\Delta \theta}{\mathrm{z}_\mathrm{max }^{\prime}}
\end{equation}
\begin{equation} \label{eq:r_relation}
r=\dfrac{z_\text{max}^{2}}{2 \Delta \theta}
\end{equation}
The designer can choose the value of $m$. In this work, the value of $m$ is taken as one and assumed to be constant for a given tool. Thus, by using \Cref{pq relation,eq:r_relation}, \Cref{lit equation} simplifies to\\
\begin{equation} \label{Simplified equation}
y=\frac{(z-q)^{2}}{2 p}-
    \dfrac{x_{i}}{\tan\left(\theta_{0}+\left( \dfrac {z}{z_\mathrm{max}}\right)^2 \Delta \theta\right)}
\end{equation}

The above \Cref{Simplified equation} is the directrix in plane $x=x_{i}$. The directrix is moved along the generatrix $g_{i}g_{i}$ forming a surface as shown in \Cref{Fig:Exp_Setup1}. The next step is to cut this surface using a profile described in the subsequent section.
\subsubsection{Profile generation}
\begin{figure}[H]
\centering
\subfigure[]{
\includegraphics[width=0.44\textwidth]{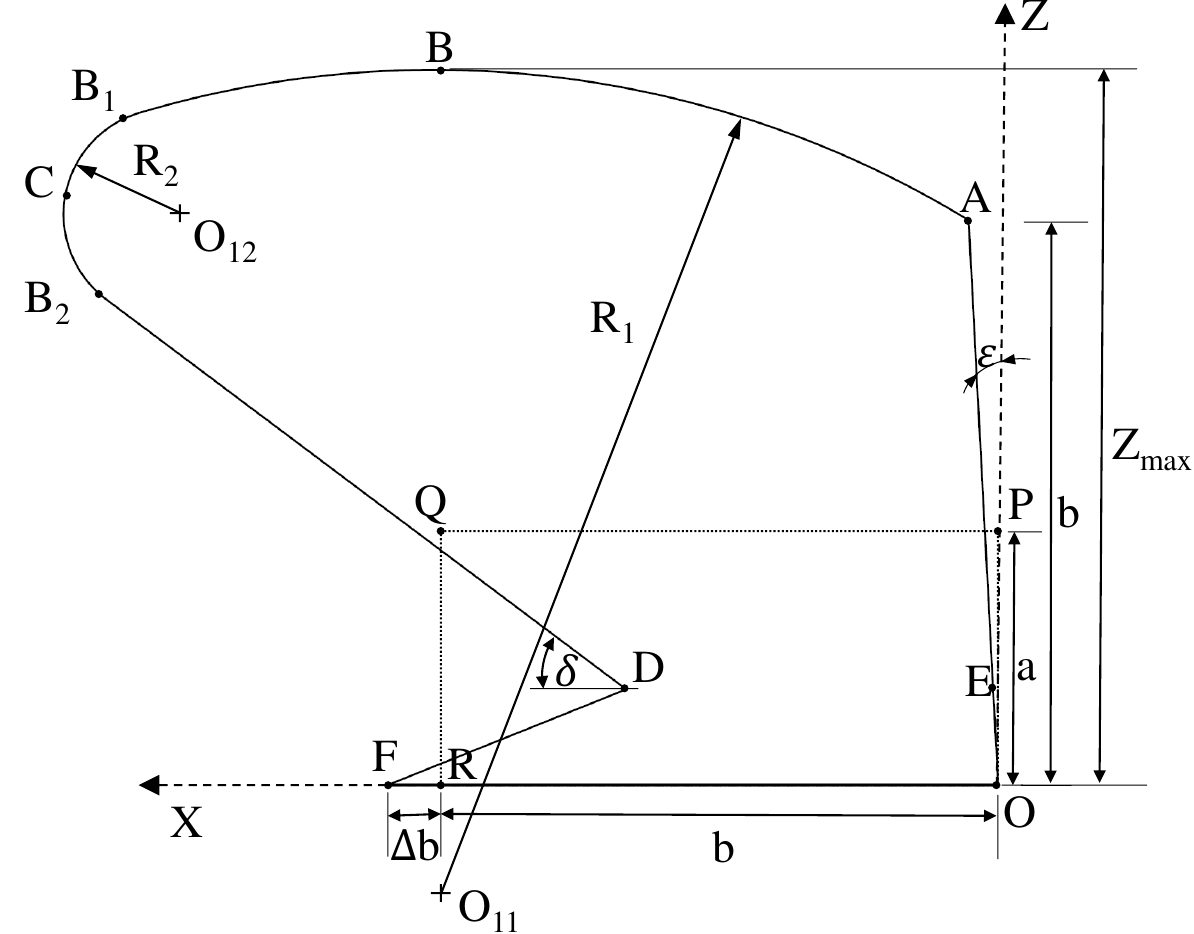}
\label{Fig:Lit_tool_profile}}%
\subfigure[]{
\includegraphics[width=0.56\textwidth]{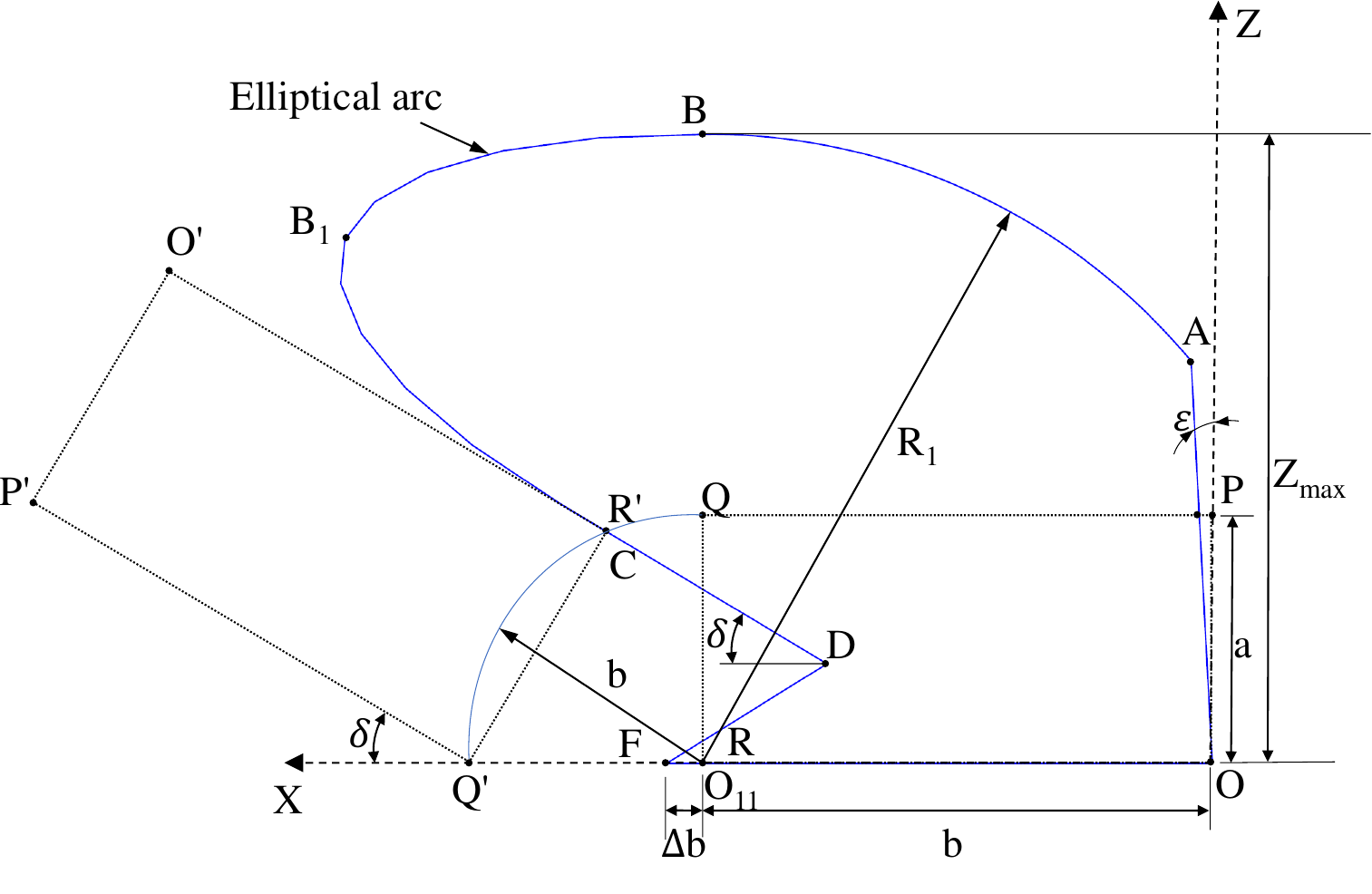}
\label{Fig:Mod_tool_profile}}%
\caption{(a) Procedure to design MB plough, and (b) Triangular wedge for defining the geometrical parameters of a passive tillage tool}
\label{Fig:2}
\end{figure}

The surface generated will be cut on a particular shape called profile, constructed on the 2D plane XOZ, as shown in \Cref{Fig:Exp_Setup1}. This profile is projected on the surface generated as described above and is trimmed to obtain the actual MB plough. The literature tool of which the profile is proposed by \citet{Ibrahmi2017b} and the modified profile are compared in this work. 
\Cref{Fig:Lit_tool_profile} shows the tool profile in the XOZ plane as proposed by \citet{Ibrahmi2017b}. The motivation behind such type of profile is to invert the soil slice, which is rectangular in shape of height a and width b about the turning point O$_{11}$. The angle between OZ and OA is the clearance angle ($\varepsilon$) chosen by the designer. The angle up to which the soil layer is to be inverted before the inversion takes place due to the gravity is given by,
\begin{equation} \label{eq: delta}
\delta = \sin^{-1}\left(\dfrac{a}{b}\right)
\end{equation} 
ABB$_{1}$ is a circular arc with centre O$_{11}$ and B$_{1}$CB$_{2}$ is a circular arc with centre O$_{12}$. OFCE represents the shape of the MB plough. \Cref{Fig:Mod_tool_profile} shows the modified profile of the MB plough for the same size of soil slice. The profile is drawn graphically, and dotted lines represent construction lines. Rectangle OPQR represents the rectangular slice of soil which is to be inverted. Compared to the original profile, two major changes were made in the modified design, i.e.,
\begin{enumerate}
    \item Since the soil slice is to be inverted at the same horizontal level, the turning point $O_{11}$ is shifted to the cutting edge on point R. OPQR and O$^\prime$P$^\prime$Q$^\prime$R$^\prime$ represent the initial and final orientation of the soil slice, respectively.
    \item To increase the surface area at the tip of the MB plough, the curve B$_{1}$B$_{2}$ is changed from circular to elliptical
\end{enumerate}
To construct the profile, the slice OPQO$_{11}$ is tilted anticlockwise about turning point O$_{11}$ such that $Q$ touches the x-axis, and the angle between the PQ and the x-axis is $\delta$. Thus, the vertex $O$ of the soil slice is at C. BB$_{1}$C is an elliptical curve, whereas AB is a circular arc of radius R$_{1}$ given by,
\begin{equation}
    R_{1}=h+\Delta h
\end{equation}
where,
\begin{equation}
    h =  \sqrt{a^2 + b^2}
\end{equation}
and $\Delta h$ is the variation of the maximum height $h$, which varies from -25 to \SI{25}{\milli\meter} and can be chosen as per the designer experience \citep{Ibrahmi2017b}. The input parameters are specified in the \Cref{tab:profile_parameters}. The influence of these parameters on plough surface geometry and tillage performance has been comprehensively analyzed in \cite{Ibrahmi2017b}. In the present work, these parameters were directly implemented to construct the plough geometry, and a detailed reproduction of their effects is omitted for brevity.

\begin{table}[ht!]
    \centering
    \caption{Tool surface profile parameters}
    \begin{tabular}{lll}
    \hline
    \hline
     \textbf{Parameters}  & \textbf{Defination} & \textbf{Values}\\
     \hline
     \hline
    $\alpha_\text{o}$     & Cutting angle of the tool & $\tan \alpha_\text{0}=\tan \gamma_\text{0} \sin \theta_\text{o}$\\
    $\beta_\text{o}$ & Lateral vertical angle of the tool & Dependent parameter\\
    $\gamma_\text{o}$ & Rake angle & \SI{30}{\degree} \\
    $\theta_\text{o}$ & Lateral directional angle & \SI{45}{\degree} \\
    $\Delta \theta$ & Change in lateral directional angle & \\ & as generatrix move up in Z direction & \SI{17}{\degree}\\
    a & Height of furrow & 150 mm\\
    b & Width of furrow & 250 mm\\
    p, q & Geometric parameters & $p = q\tan \alpha_\text{o}$, q = 250 mm\\
    \hline
    \hline
    \end{tabular}
    \label{tab:profile_parameters}
\end{table}

The calculated surface area was found to be \SI{101.859}{\centi\meter\squared} and \SI{108.568}{\centi\meter\squared} for literature and modified MB plough surface, respectively. The changes in the modified profile result in an increase in surface area of 6.6\% as compared with the literature tool.

Four design parameters significantly affect the shape of the MB plough, i.e. $\Delta$, $\theta$, $\theta_{0}$, $\gamma_{0}$ and $q$, are described in supplementary.
\subsection{Calculation of Inversion Index}
\begin{figure}[H]
\begin{center}
\subfigure[]{
\includegraphics[width=0.6\textwidth]{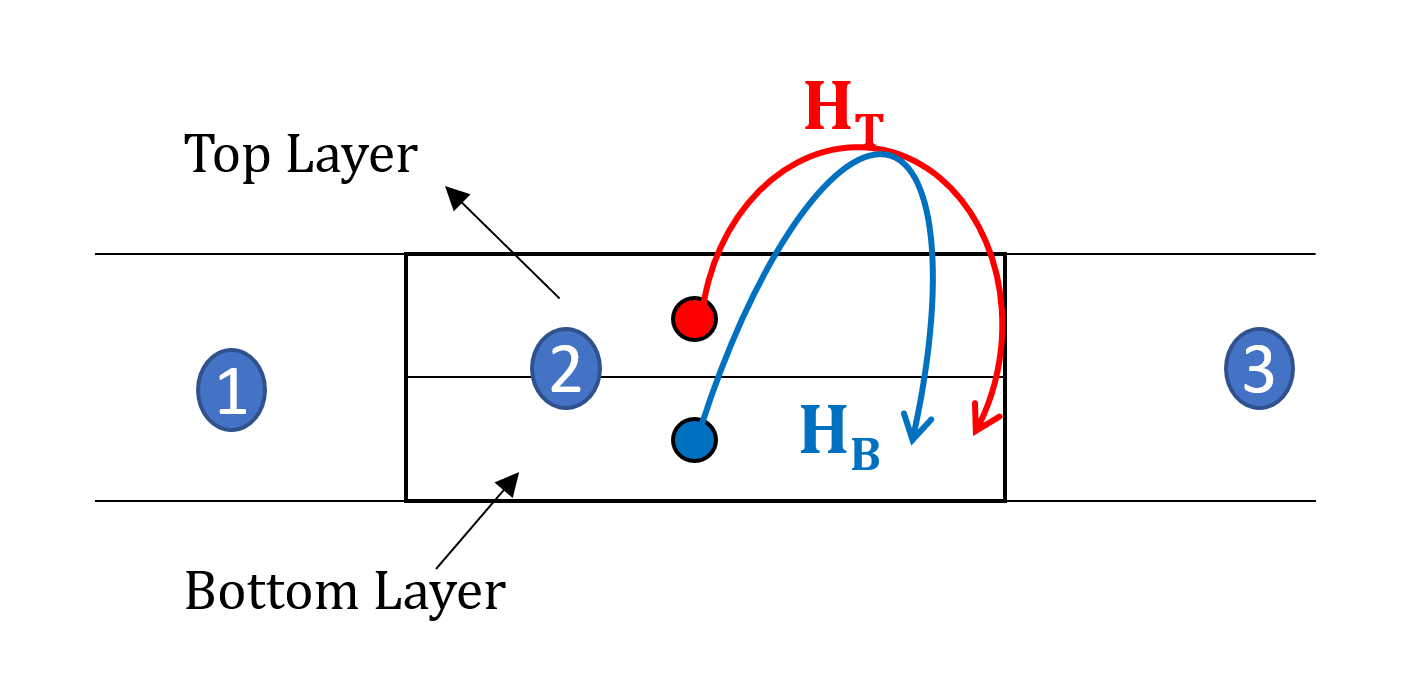}
\label{fig:soil_displacement_horizontal}}
\subfigure[]{
\includegraphics[width=0.32\textwidth]{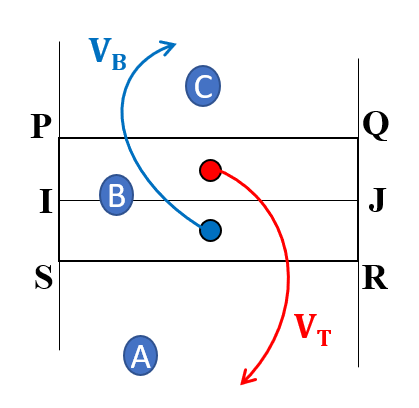}
\label{fig:soil_displacement_vertical}}
\caption{Soil layer displacement in (a) horizontal, and (b) vertical directions. The working direction of the moldboard plough is into the plane of the paper.}
\label{fig:soil_displacement}
\end{center}
\end{figure}

\begin{figure}[htbp]
\begin{center}
\subfigure[]{
    \includegraphics[width=1\textwidth]{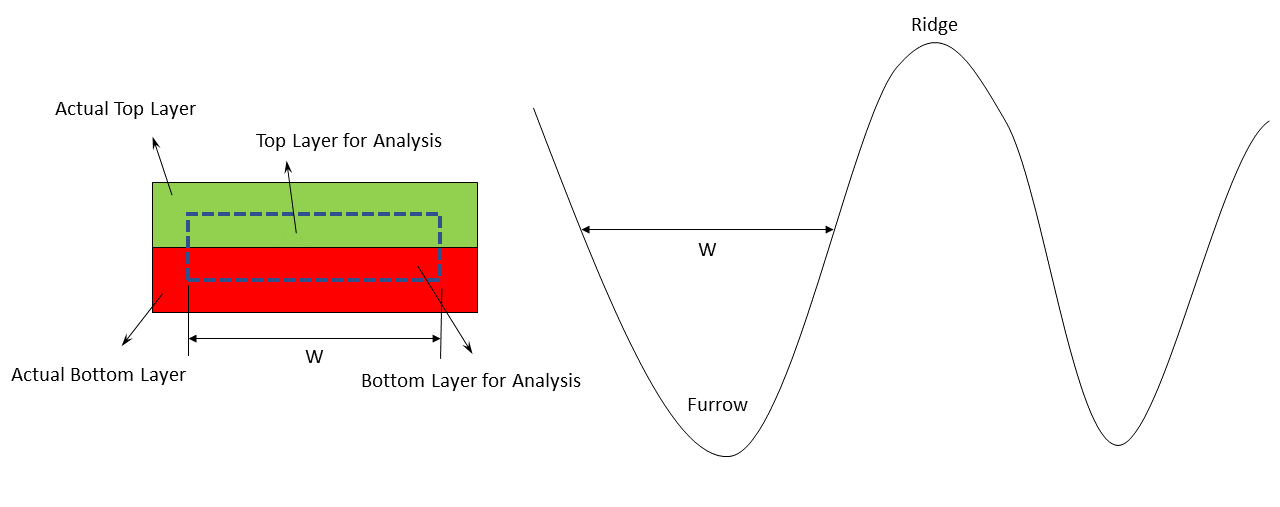}
    \label{fig: width and layer thickness}}
\subfigure[]{
\includegraphics[width=0.75\textwidth]{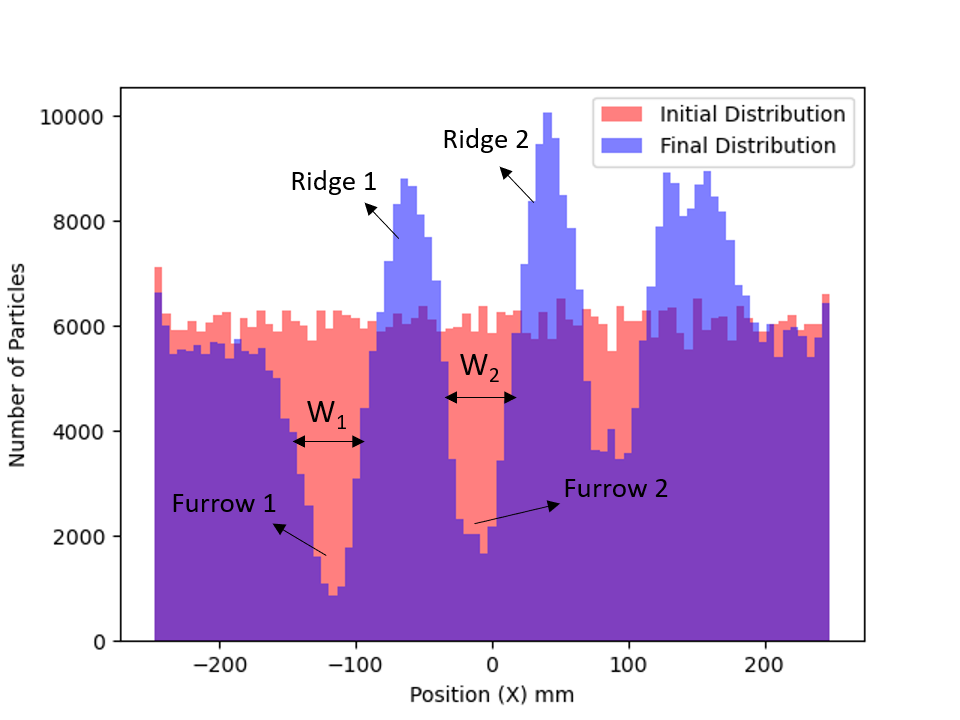}
    \label{fig: furrow and ridge profile}}
    \caption{(a)Width and Layer thickness selection for analysis, and (b) Furrow and ridge formed after ploughing}
 \end{center}   
\end{figure}

In this paper, a new method has been proposed to calculate the inversion efficiency of the plough. In the past work, the efficiency of the plough is calculated by soil disturbance by studying the furrow shape formed \citep{Ibrahmi2015}. Scanning is performed before and after the experiment to calculate the width and surface area of the cut soil. This procedure only gives the idea about the amount of soil removed from the target position, but does not provide any information about how the soil is inverted. Therefore, an index is needed to predict the horizontal and vertical movements of the soil layers. In this work, a new method to determine the inversion efficiency of the tool, which is predicted by a parameter called the inversion index, has been proposed. In this method, the target slice PQRS (see \Cref{fig:soil_displacement}) of the soil is divided into two vertical layers, that is, the top and bottom layers represented by PQJI and IJRS, respectively. 
As the furrow formed is of variable width, the width of the soil taken for the analysis of a particular furrow is equal to the mean width of the respective furrow as shown in \Cref{fig: width and layer thickness}. To calculate the inversion index, the movement of the soil layers horizontally and vertically needs to be studied. Only half of the top layer and bottom layer are considered (see \Cref{fig: width and layer thickness}) to calculate the inversion efficiency because it allows the tracing of the particles more efficiently below the bottom as well as the top layer of the soil. There is no space below the bottom layer where we could trace the soil particles if the entire soil layer is taken into consideration.
The width of the target slice is equal to the mean width of furrow formed $W_{1}$ and $W_{2}$ for furrow 1 and 2, respectively, as shown in \Cref{fig: furrow and ridge profile}. To trace the particles, the entire region is divided into 3 regions horizontally (regions 1, 2, \& 3) and vertically (regions A, B, \& C) as shown in \Cref{fig:soil_displacement}. Region 1, 2 and 3 are the regions on the left, middle and right-hand side of the target slice PQRS, respectively. In a similar way, regions A, B and C are the regions on the bottom, middle and top of the target slice.
\Cref{fig: furrow and ridge profile} shows the particles distribution along the horizontal direction before and after ploughing. The mould-board ploughs are attached in series on a particular angle in such a way that the front plough makes the empty space by removing and inverting the soil to the right. The empty space created by this plough is filled with soil which is inverted by the following back plough. The simulation is done with the three ploughs, out of which the furrow and ridge formed by the two ploughs has been studied as shown in \Cref{fig: furrow and ridge profile}. Inversion index is calculated based on the mass fraction transferred from a particular region to other. The target soil slice PQRS is traced by its position in vertical and horizontal direction as seen in \Cref{fig:soil_displacement}. Among all the combinations of mass distributions only four mass fractions are considered and summed up i.e, $H_\text{T}$, $H_\text{B}$, $V_\text{T}$ and $V_\text{B}$. $H_\text{T}$ and $H_\text{B}$ represents the lateral throw whereas $V_\text{T}$ and $V_\text{B}$ represents the vertical throw of the soil. It is expected that for a good inversion of soil the top layer should be buried below in the same region horizontally while the bottom layer is spread over the ridge. The higher inversion index represents higher inversion efficiency of tool. 
Thus, the overall inversion index, \textbf{$I$} is given as, 
\begin{equation} \label{eq: inversion index}
I  = H_{T} + H_{B} + V_{T} + V_{B}
\end{equation}
Let us consider the notation $M_{T2f}$ represents mass of top layer in region 2 after ploughing, hence $H_{T}$, $H_{B}$, $V_{T}$ and $V_{B}$ can be calculated as,
\begin{align} \label{eq: mass_fraction_cal}
H_{T} = M_{T2f}/M_{T2i}\\
H_{B} = M_{B3f}/M_{B2i}\\
V_{T} = M_{TAf}/M_{TBi}\\ 
V_{B} = M_{BCf}/M_{BBi}
\end{align}
Further details and case studies are provided in the supplementary material.
\subsection{Simulation setup}
To simulate the ploughing process using the DEM, it is essential to accurately define the interactions between the soil particles and the soil-tool interface. These interactions are represented through contact models, which govern the forces and behaviours during contact. This study uses the Hertz-Mindlin contact model combined with the linear cohesion model to represent the soil. This approach accounts for the cohesive behaviour of the soil particles, providing a more realistic simulation of soil mechanics during ploughing. The equations used in the simulations are detailed in the supplementary material.

\begin{figure}[H]
    \centering
    \includegraphics[width=1\textwidth]{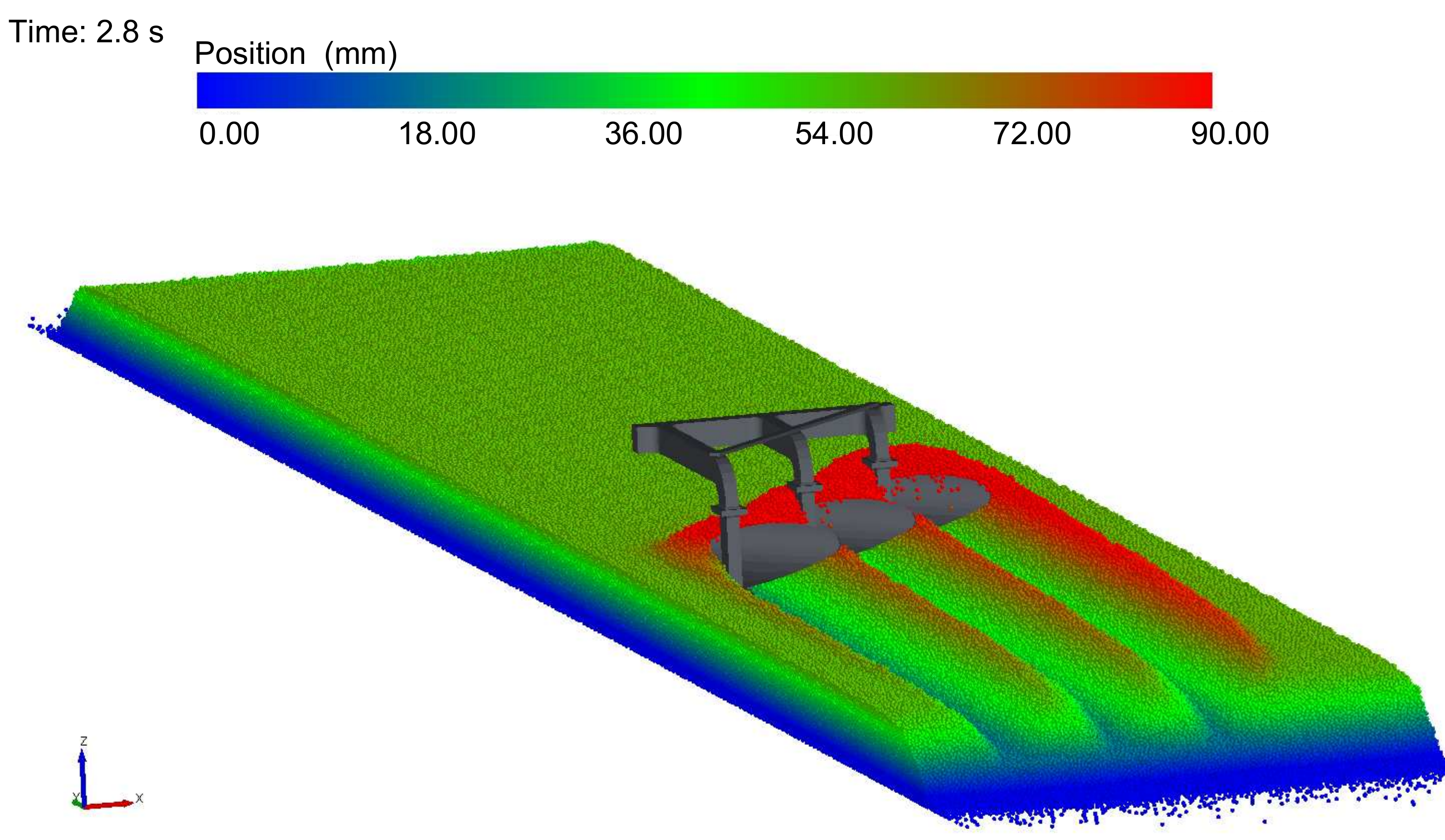}
    \caption{3D view of ploughing simulation}
    \label{fig: 3D_view_of_simulation}
\end{figure}

\begin{figure}[H]
\centering
\subfigure[]{
\includegraphics[width=0.9\textwidth]{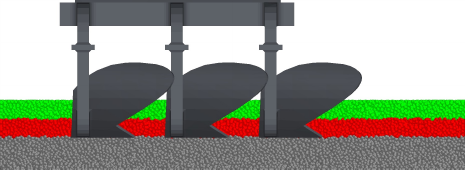}
\label{fig:initial config}}
\subfigure[]{
\includegraphics[width=0.9\textwidth]{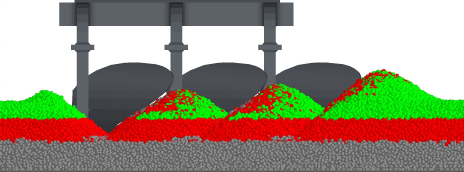}
\label{fig:final config}}%
\caption{ (a) Initial and (b) final configuration of ploughing}
\label{fig:configuration}
\end{figure}

The simulation is performed at a reduced scale of $1/4^\text{th}$ of the actual ploughing process to reduce computational time, while maintaining full-scale particle properties, contact parameters, and tool material properties to preserve the accuracy of stress distribution and wear predictions. \Cref{fig: 3D_view_of_simulation} shows the 3D view of the simulation performed in EDEM software \citep{EDEM}, an Altair DEM tool. Firstly, a \SI{0.1}{\meter} thick soil bed of \SI{3}{\meter} in length and a width of \SI{0.5}{\meter} is prepared. \Cref{fig:initial config} shows the ploughing tool's initial position. The MB plough assembly is run through a length of \SI{3}{\meter} through the soil bed, which results in the ridge and furrow profile of the soil bed as shown in \Cref{fig:final config}. Additional details of the setup are provided in the supplementary material. 
%

\subsection{Model validation} \label{subsec: angle of repose}
\begin{figure}[H]
\centering
\subfigure[]{
\includegraphics[width=0.8
\textwidth]{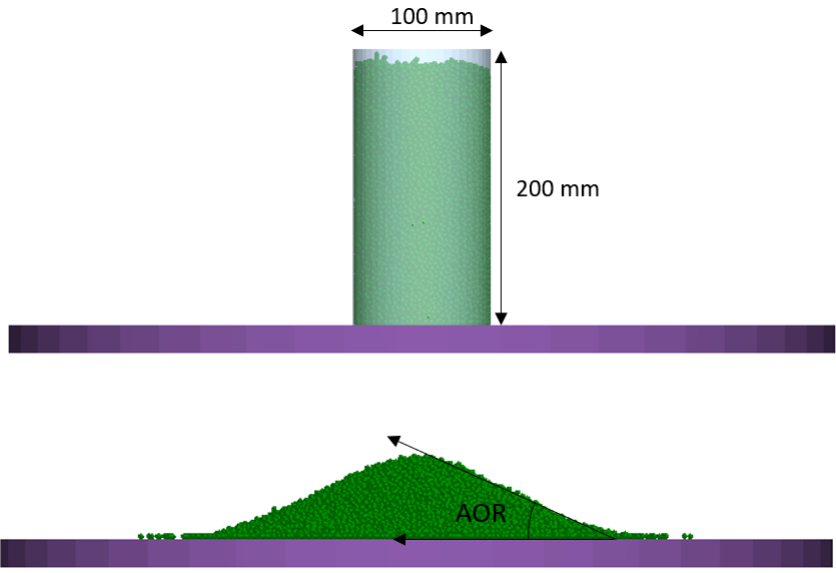}
\label{Fig:AOR_setup}}
\subfigure[]{
\includegraphics[width=0.8\textwidth]{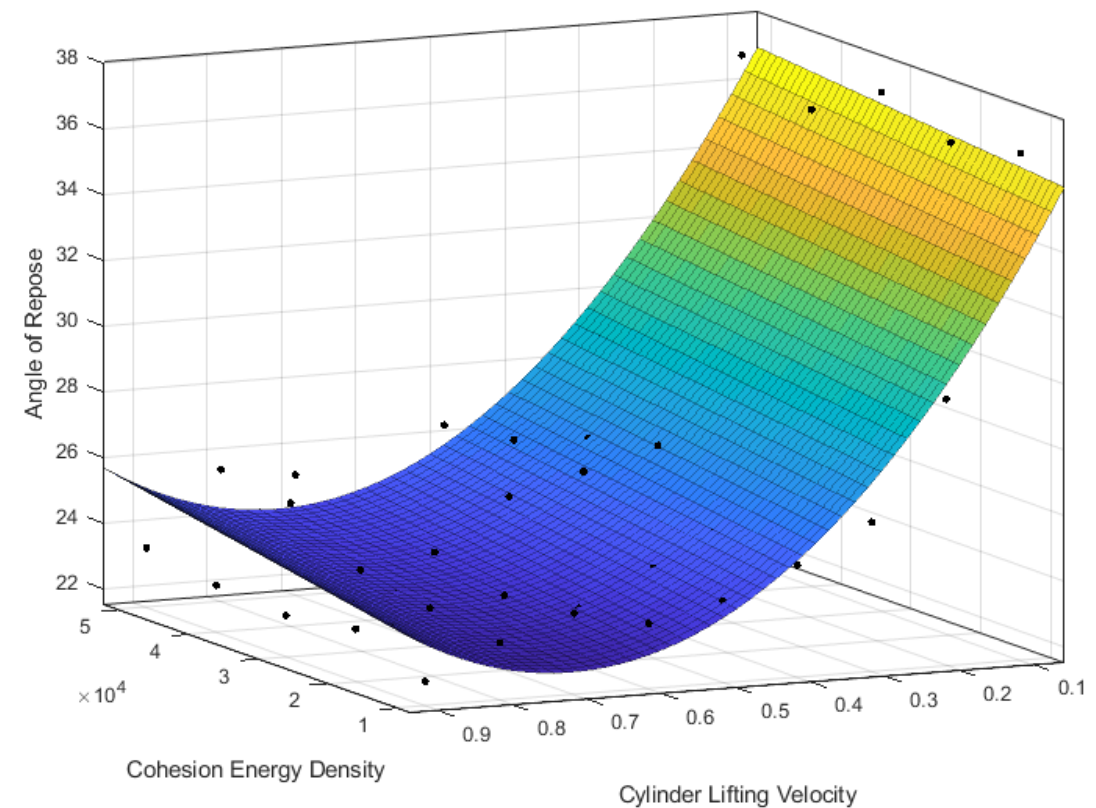}
\label{Fig:AOR_plot}}%
\caption{(a) Angle of Repose setup, and (b) Surface plot of variations of AOR vs CED and lifting velocity}
\label{Fig: Validation}
\end{figure}

The angle of repose can be defined as the angle formed by the heap of particles with the horizontal, as shown in \Cref{Fig:AOR_setup}. A simulation study is performed on the effect of cohesion energy density and lifting velocity of the cylinder on the angle of repose. Soil particles are filled up to the top of the cylinder with the dimensions, as shown in \ref{Fig:AOR_setup}. The cylinder is lifted with a constant velocity ranging from \SI{0.1}{\meter\per\second} to \SI{0.9}{\meter\per\second}, Cohesion Energy Density ($\Gamma$) is varied from 10,000 to \SI{50000}{\joule\per\meter\cubed}, and the angle of repose is measured. \Cref{Fig:AOR_plot} shows the surface plot of the variation of AOR with the $\Gamma$ and lifting velocity. It is observed that the lifting velocity has a much higher influence on AOR than $\Gamma$. The parameters used in the simulations are listed in \Cref{tab:DEM_parameter}
\begin{table}[!ht]
\centering\caption{Parameters used for DEM Simulation}
\begin{threeparttable}
\begin{tabular}{lll}
\hline
\hline
\textbf{Parameters} & \textbf{Value} & \textbf{Source} \\
\hline
\hline
$R$   & \SI{2}{\milli\meter}& Selected\\
$\rho_\text{s}$ & \SI{2600}{\kilo\gram\per\metre\cubed} & \citet{Barr2018}\\
$\rho_\text{p}$ & \SI{7850}{\kilo\gram\per\meter\cubed} & \citet{Barr2018}\\
$v_\text{s}$ & $0.3$ & \citet{Barr2018}\\
$v_\text{p}$ & $0.3$ &\citet{Barr2018} \\
$E_\text{s}$ & \SI{0.1}{\giga\pascal} & \citet{Lommen2014}\\
$E_\text{p}$  & \SI{205.4}{\giga\pascal} & Structural steel\\
$\Gamma_\text{ss}$ &\SI{30000}{\joule\per\meter\cubed} & \citet{Ucgul2017}\\
$\Gamma_\text{sp}$ & \SI{0}{\joule\per\meter\cubed} & Selected\\
$e_\text{ss}$ & $0.5$ & Selected\\
$e_\text{sp}$  & $0.5$ & Selected (Calibrated, AOR = \SI{28.5}{\degree})\\
$\mu_\text{ss}$  & $0.5$ & Selected (Calibrated, AOR = \SI{28.5}{\degree})\\
$\mu_\text{sp}$ & $0.5$ & \citet{Ucgul2017}\\
$\mu_{R_\text{ss}}$ & $0.2$ & Selected (Calibrated, AOR = \SI{28.5}{\degree}) \\
$\mu_{R_\text{sp}}$  & $0.2$  & Selected\\
\hline
\hline
\end{tabular}
\label{tab:DEM_parameter}
  \begin{tablenotes}
      \small
      \item Note: s denotes soil, p denotes plough.
\end{tablenotes}
\end{threeparttable}
\end{table}
\section{Results and Discussions} \label{sec: Results}
\subsection{Forces on MB ploughs} \label{sec:forces on MB plough}
The overall soil inversion process is divided into four operations to be performed on any mouldboard surface: cutting, lifting, inverting, and throwing the soil. A significant proportion of draft, vertical, and lateral force is associated with the cutting, lifting, and throwing processes. While, the shape and curvature of the mouldboard surface influence the inversion of the soil. The forces exerted by the soil particles in three directions named lateral, draft, and vertical downward forces for different velocities of the plough (0.5, 0.8, and \SI{1}{\meter\per\second}) are presented here.

\subsubsection{Draft force ($F_\text{y}$), lateral force ($F_\text{x}$), and vertical downward force ($F_\text{z}$)}

\begin{figure}[H]
\centering
\subfigure[]{
\includegraphics[width=0.5\textwidth]{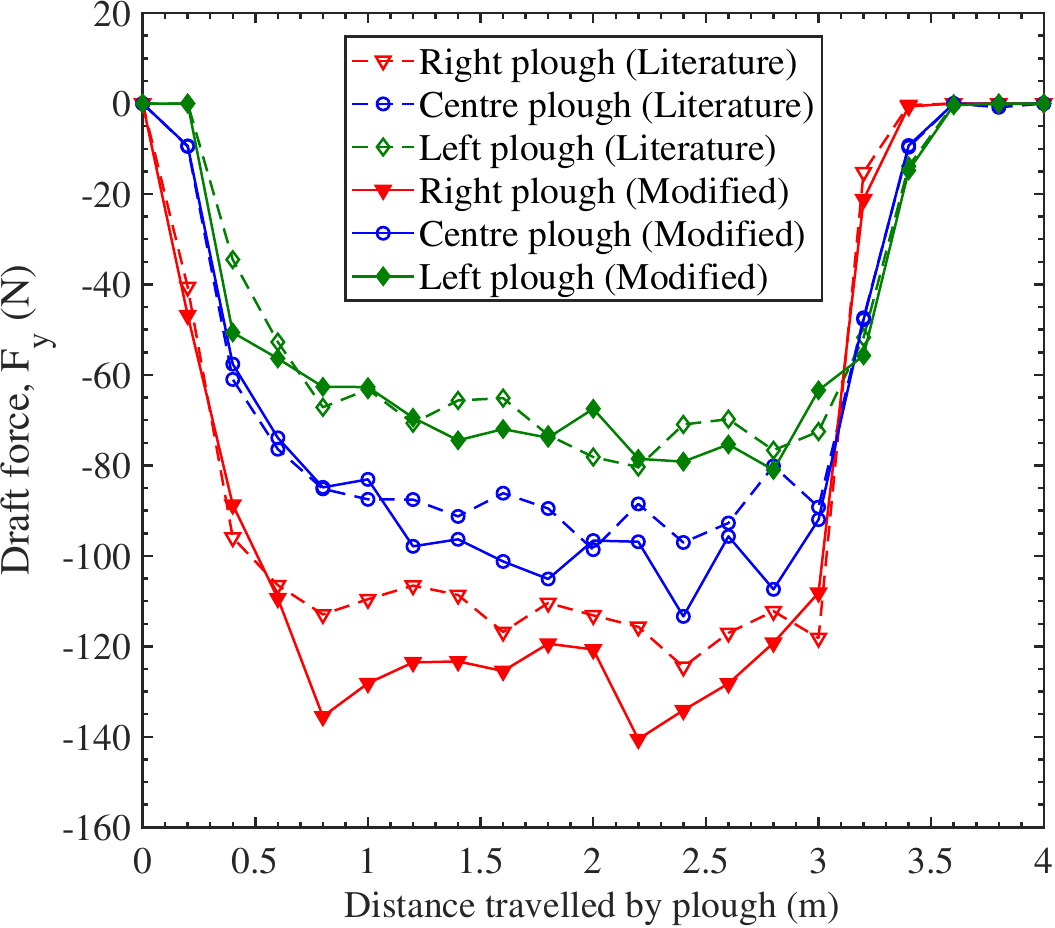}
\label{Fig:Forces on plough}}%
\subfigure[]{
\includegraphics[width=0.465\textwidth]{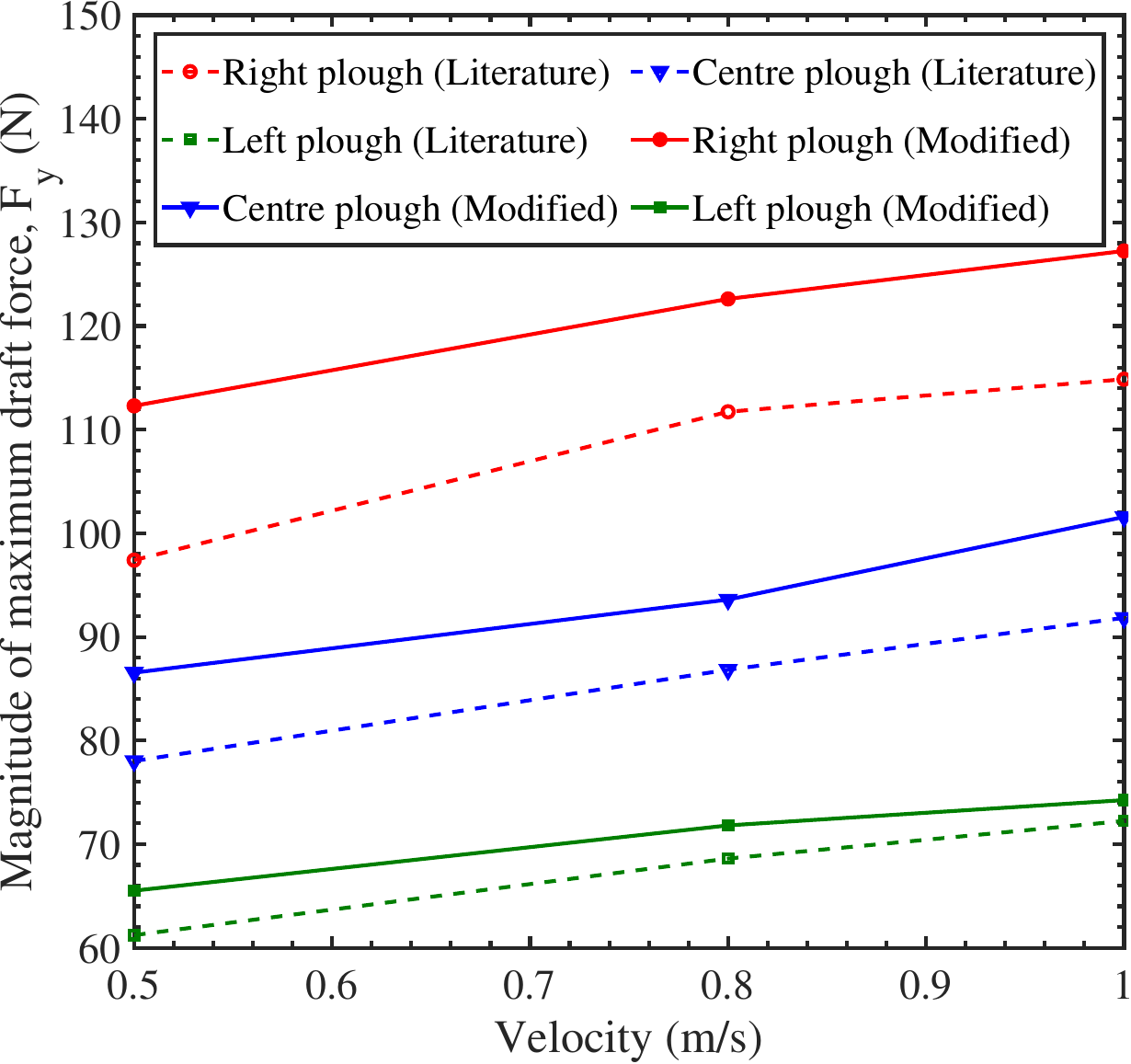}
\label{Fig:ForcevsVelocity}}
\caption{(a) Draft forces on MB plough for $v$ = 1 m/s, (b) variation of draft force with velocity on three ploughs}
\label{Fig:Draft forces on MB plough}
\end{figure}

Draft force ($F_\text{y}$) exerted by the soil particles on the mouldboard surface plays a vital role in fuel consumption by the tractor while ploughing. To study the effect of the position of the plough on the draft force, the variation of the magnitude of draft force with respect to the direction travelled is studied. It is clear from \Cref{Fig:Forces on plough} that the right plough experiences more draft among the three ploughs because of the unavailability of the vacant furrow space on the right side of the right plough. Hence, the quantity of soil lifted by the right plough is the highest among the three ploughs. In actual field operation, several MB ploughs are attached in series at an inclination to the tractor arm. The centre and leftmost ploughs resemble the ploughs in series at a particular inclination. Initially, the mass rate of soil coming in contact with the plough starts increasing for a certain distance, becomes stable afterwards,. Then, it again starts decreasing while leaving the soil bed. The average forces are calculated for the middle region of the total travel. \Cref{Tab:Forces on MB ploughs} shows the average lateral, draft and vertical compressive force on the plough surface at velocities of 0.5, 0.8 and \SI{1}{\meter\per\second}. Maximum draft force on the three ploughs increases with an increase in velocity from 0.5 to \SI{1}{\meter\per\second}, as shown in \Cref{Fig:ForcevsVelocity}. The draft force required to drive the plough is compared for literature and modified design. It is observed that there is a 15.28\%, 10.60\% and 7\% relative increase in draft force for right, centre and left ploughs for modified design as compared with the literature, as shown in \Cref{Tab:Forces on MB ploughs}. The draft force has been observed to be 25 to 30 times greater than the lateral force, as discussed below.

The simulation results indicate that the improved plough design, while enhancing soil inversion, results in a marginal increase in draft force compared to the baseline. In practical field operations, increased draft force directly translates to higher fuel consumption and energy demand, which may elevate operational costs. Moreover, elevated draft forces can impose greater mechanical loads on the tractor–implement system, potentially leading to increased wear and tear, reduced equipment lifespan, and lower overall field efficiency. These implications underscore the importance of a balanced design approach that considers both tillage effectiveness and energy efficiency. Future design iterations may explore structural or material optimizations to mitigate these forces without compromising functional performance.

Lateral force ($F_\text{x}$) refers to the side thrust experienced by the ploughing surface along the x-direction. In this work, the lateral force on the right, centre, and left plough is examined for various velocities of the plough like \SI{0.5}{\meter\per\second}, \SI{0.8}{\meter\per\second}, and \SI{1}{\meter\per\second}. This force is the lowest for the centre plough among the three ploughs for all three velocities, as shown in \Cref{Tab:Forces on MB ploughs}. Additionally, the results indicate the correlation between the velocity and lateral force. Specifically, with an increase in the velocity from \SI{0.5}{\meter\per\second} to \SI{1}{\meter\per\second}, there is a slight increase in the lateral force exerted on the ploughing surface. This finding suggests that higher velocities may lead to greater side thrust forces, potentially impacting the stability and wear of the plough.

When the plough lifts soil particles, they exert a downward force vertically on the plough surface, the magnitude of which depends on the mass of the soil lifted by the plough. From the \Cref{Tab:Forces on MB ploughs}, it is clear that the centre plough experiences the highest vertical downward force among the three ploughs for both the MB plough profile (literature and modified). Also, when $F_\text{z}$ is compared for literature with a modified design, it shows that soil particles impart more force in a downward direction for the modified design. Hence, it can be concluded that more soil particles are lifted, and therefore, the modified plough is more stable than the literature one as it has more grip on the ground, which allows the plough to excavate the soil efficiently. These observations are significant for optimizing plough design and operational parameters.
\begin{table}[H]
\caption{Average Draft, Lateral and Vertical force on ploughs (\textit{PR-Right Plough, PC-Centre Plough, PL-Left Plough})}
\begin{tabular}{llccccccccc}
\hline
\hline
\multirow{2}{*}{Velocity} &
  \multicolumn{1}{c}{\multirow{2}{*}{Tool}} &
  \multicolumn{3}{c}{Lateral ($F_{x}$, N)} &
  \multicolumn{3}{c}{Draft ($F_{y}$, N)} &
  \multicolumn{3}{c}{Vertical ($F_{z}$, N)} \\
  \cmidrule(lr){3-5}\cmidrule(lr){6-8}\cmidrule(lr){9-11}
 & \multicolumn{1}{c}{} & PR & PC & PL & PR & PC & PL & PR & PC & PL \\
 \hline
 \hline
\multirow{2}{*}{0.5 m/s} & Lit & -9.96 & -2.97 & -19.91 & -97.41 & -78.02 & -61.22    & -19.11 & -23.53 & -21.38    \\
& Mod   & -11.28  & -2.89 & -21.56  & -112.3 & -86.55  & -65.51 & -22.9  & -26.78   & -23.45 \\ 
 \midrule
 \multirow{2}{*}{0.8 m/s} & Lit & -10.67 & -5.79  & -24.1  & -111.73   & -86.83    & -68.61    & -18.31    & -22.72    & -21.27    \\
& Mod & -13.78  & -3.85 & -23.54 & -122.62 & -93.6  & -71.81 & -22.19    & -30.7     & -23.8 \\ 
\midrule
\multirow{2}{*}{1.0 m/s} & Lit           & -10.61      & -3.69     & -24.58    & -114.87   & -91.84    & -72.24    & -18.51    & -24.76    & -20.73    \\
                               & Mod             & -15.11      & -8.50     & -25.87    & -127.25   & -101.58   & -74.25    & -30.49    & -24.99    & -20.80   \\
\hline
\hline
\end{tabular}
    \label{Tab:Forces on MB ploughs}
\end{table}
%
\subsection{Inversion Index} \label{sec:Inversion index}
\begin{figure}[H]
\centering
\subfigure[]{
\includegraphics[width=0.49\textwidth]{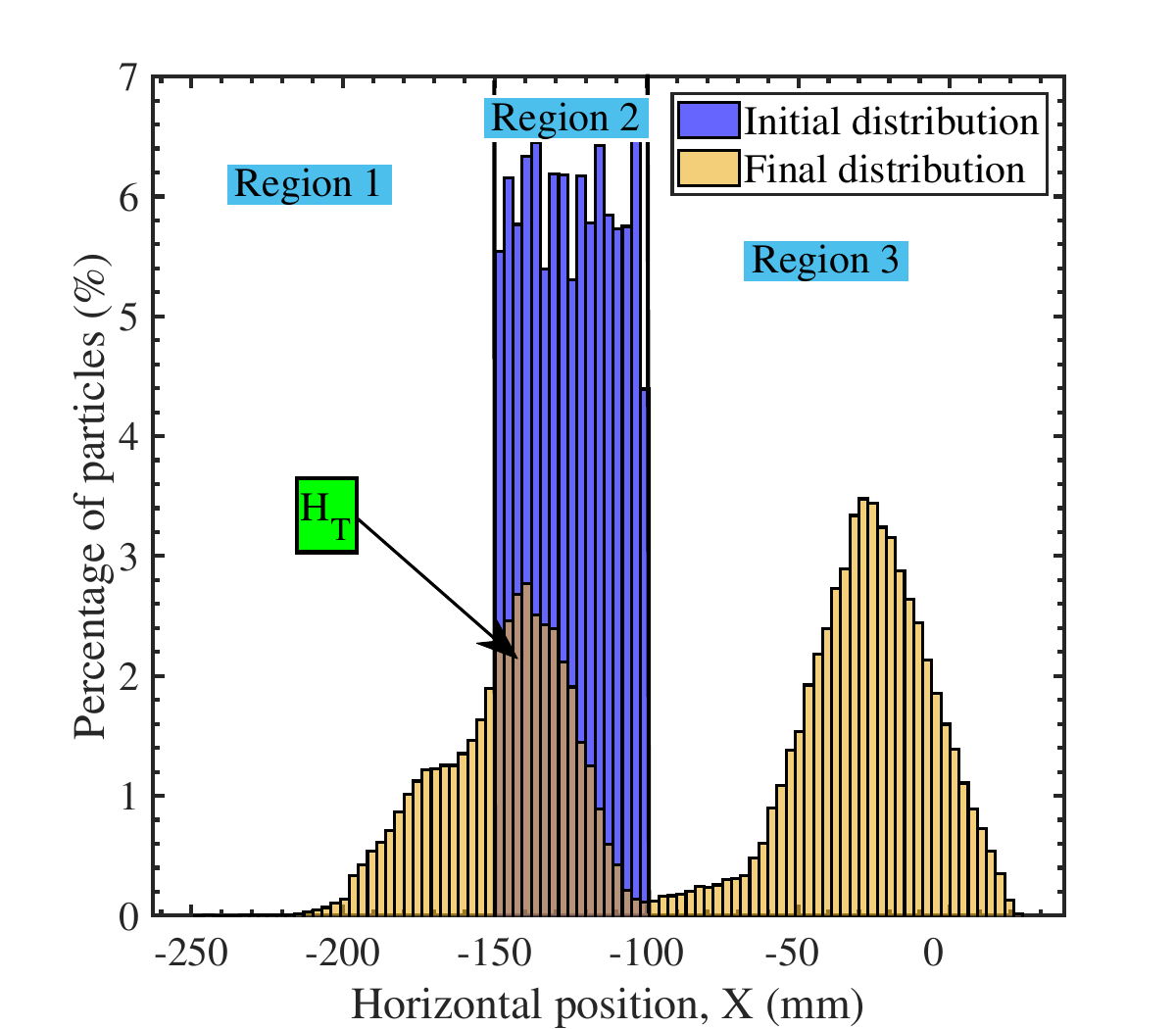}
\label{Fig:Top_Horizontal_distribution}}%
\subfigure[]{
\includegraphics[width=0.49\textwidth]{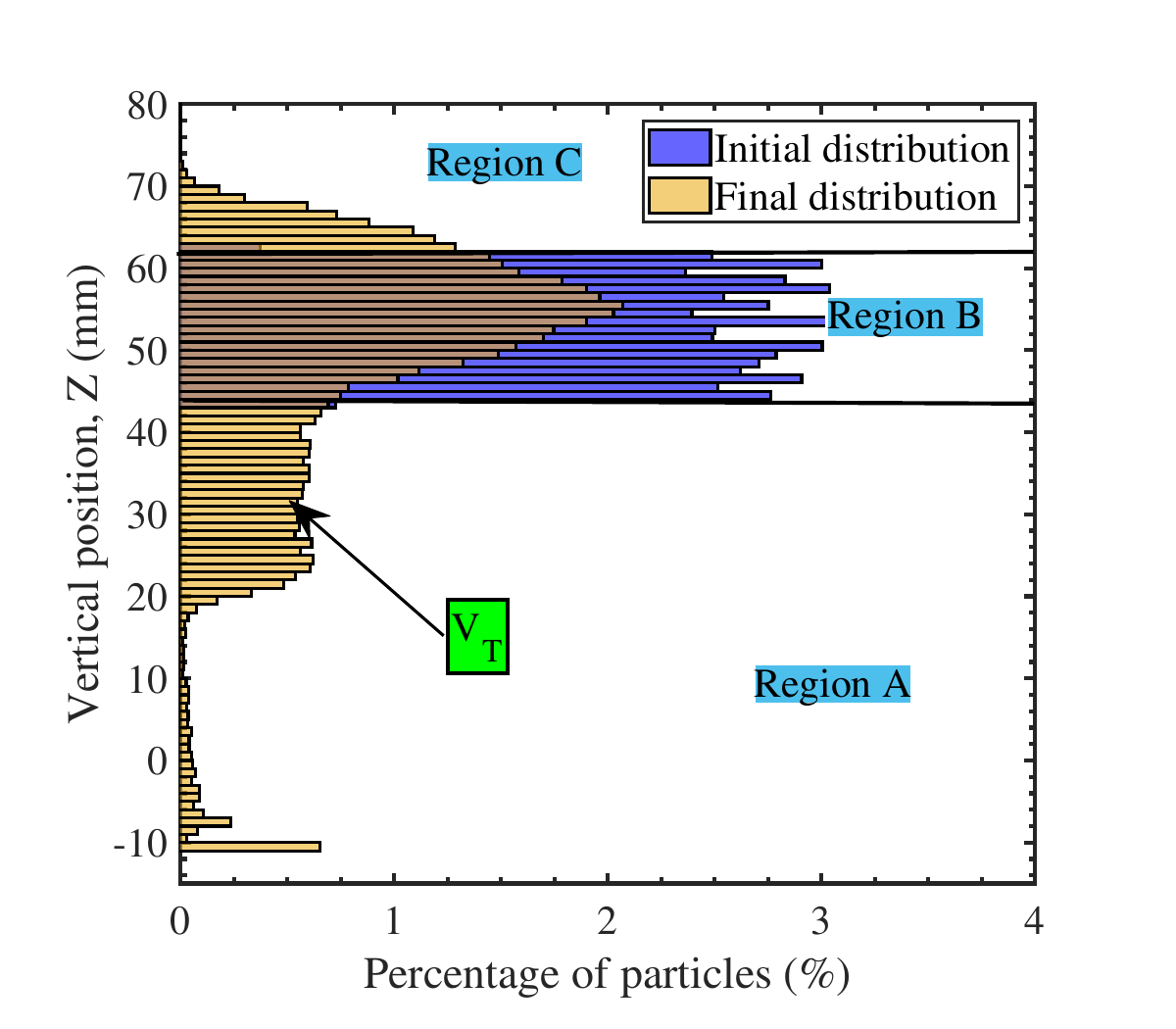}
\label{Fig:Top_Verical_distribution}}
\subfigure[]{
\includegraphics[width=0.49\textwidth]{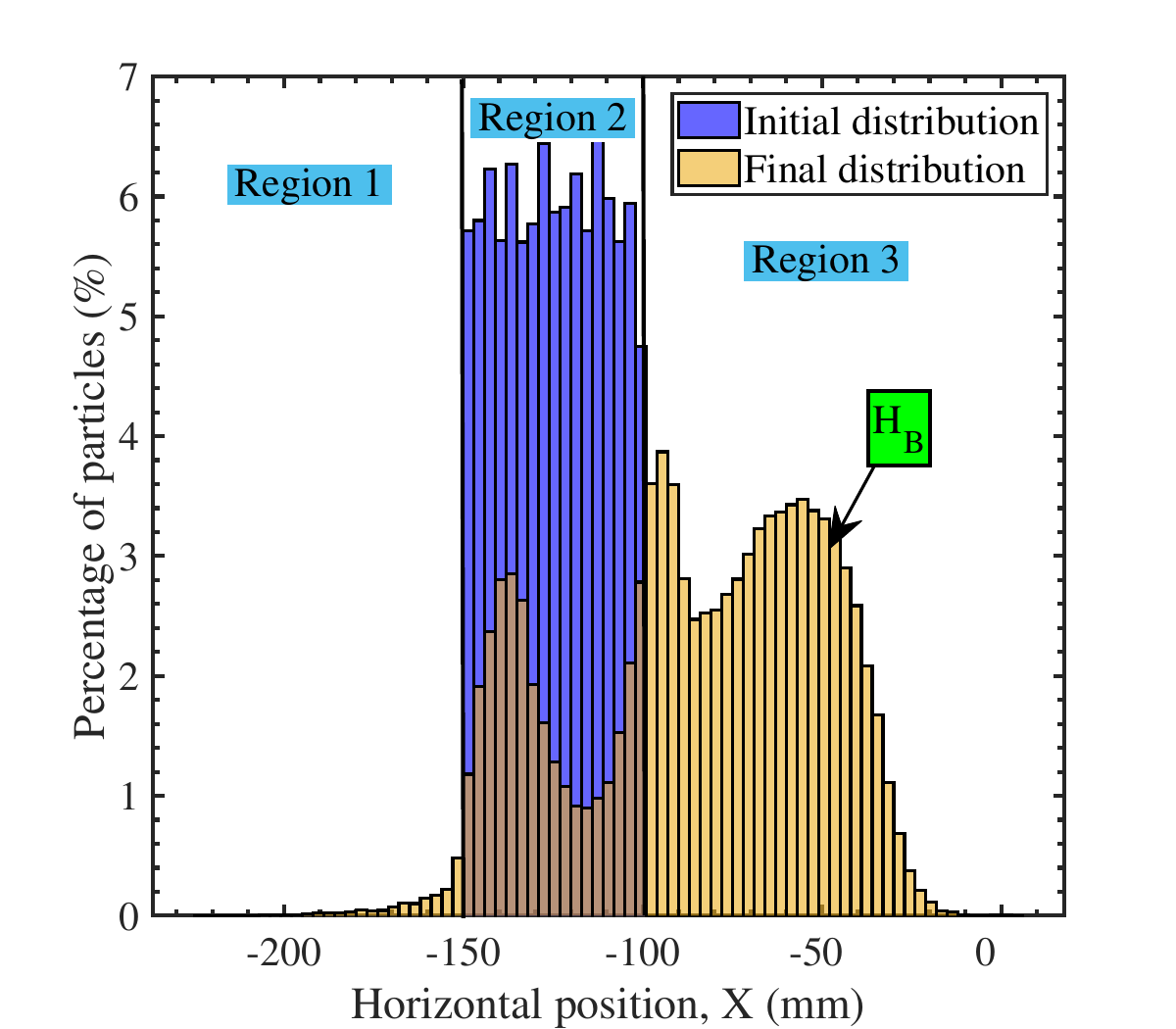}
\label{Fig:Bottom_Horizontal_distribution}}%
\subfigure[]{
\includegraphics[width=0.49\textwidth]{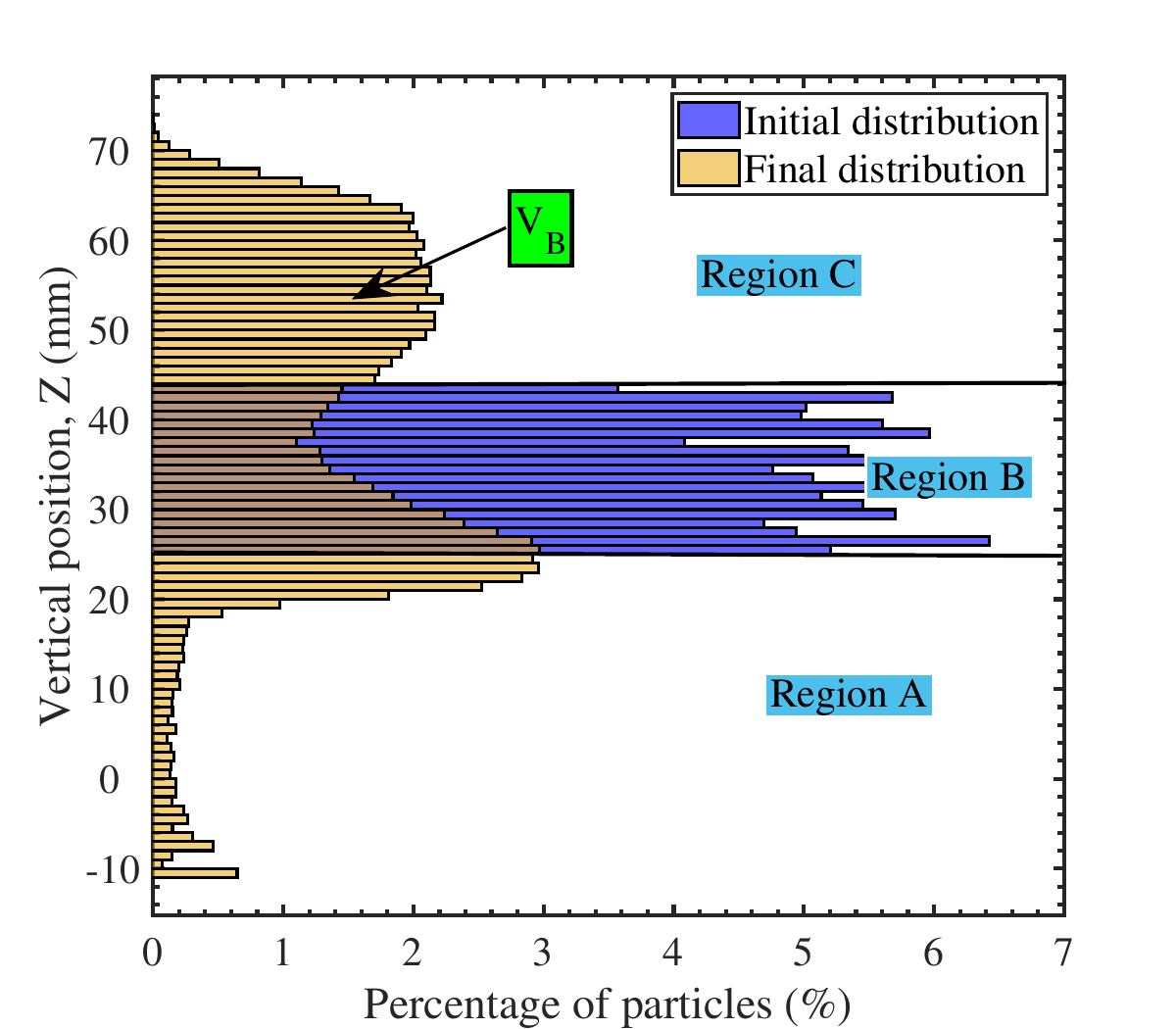}
\label{Fig:Bottom_Vertical_distribution}}
\caption{Distribution of soil particles in the (a) top and (b) bottom layers}
\label{Fig:Histograms_top_bottom_layer}
\end{figure}

\begin{figure}[H]
\centering
\subfigure[]{
\includegraphics[width=0.33\textwidth]{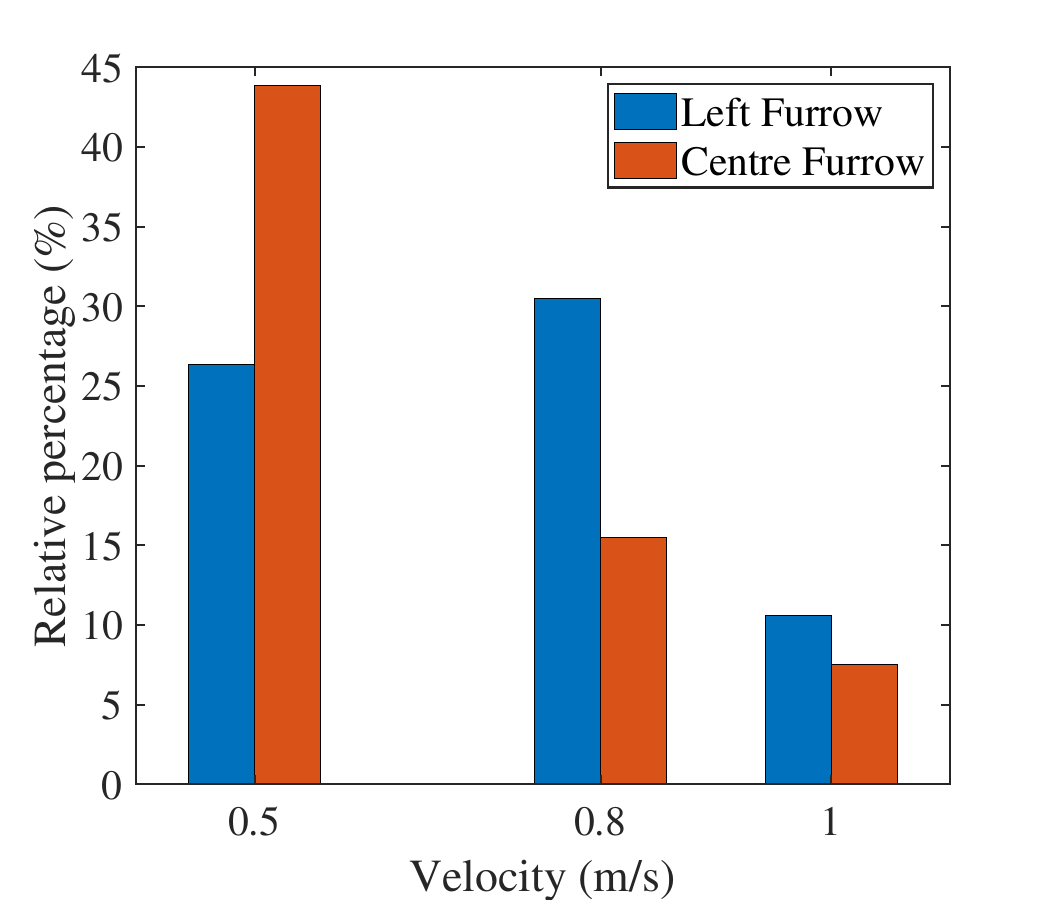}
\label{Fig:Relative increase in inversion index horizontal}}%
\subfigure[]{
\includegraphics[width=0.33\textwidth]{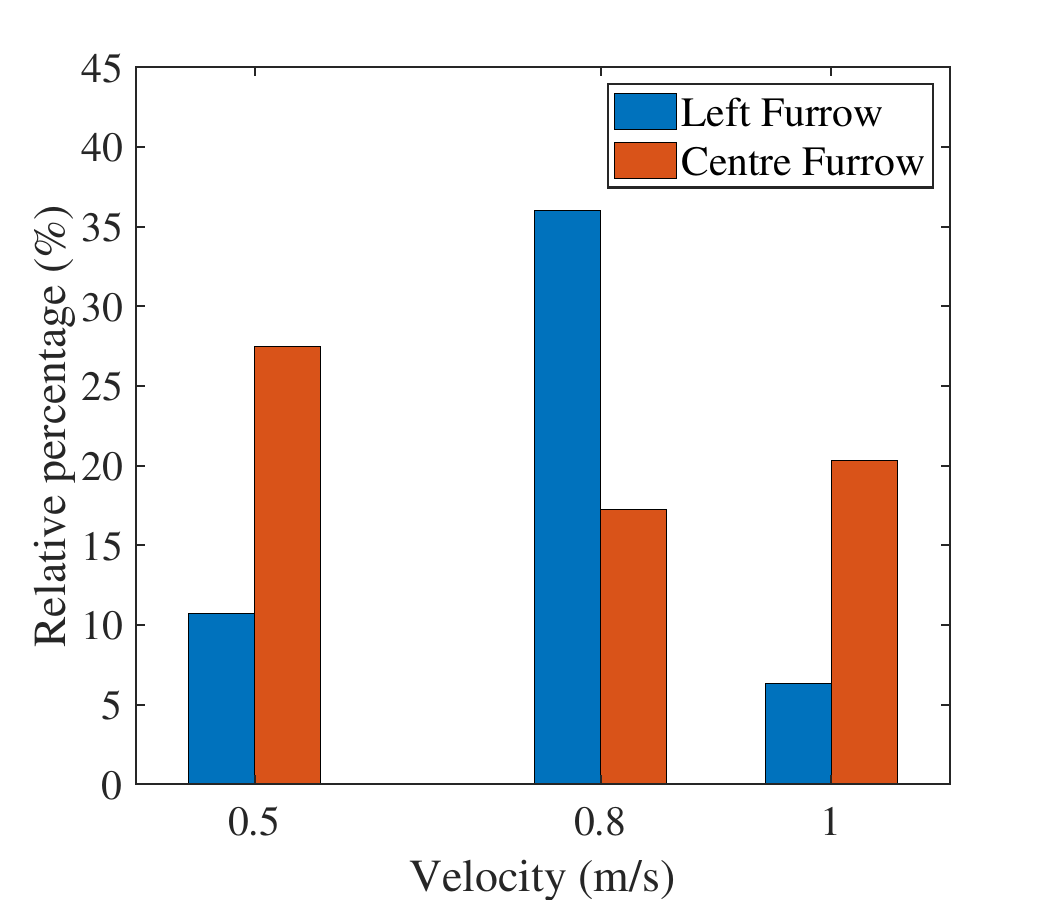}
\label{Fig:Relative increase in inversion index vertical}}%
\subfigure[]{
\includegraphics[width=0.33\textwidth]{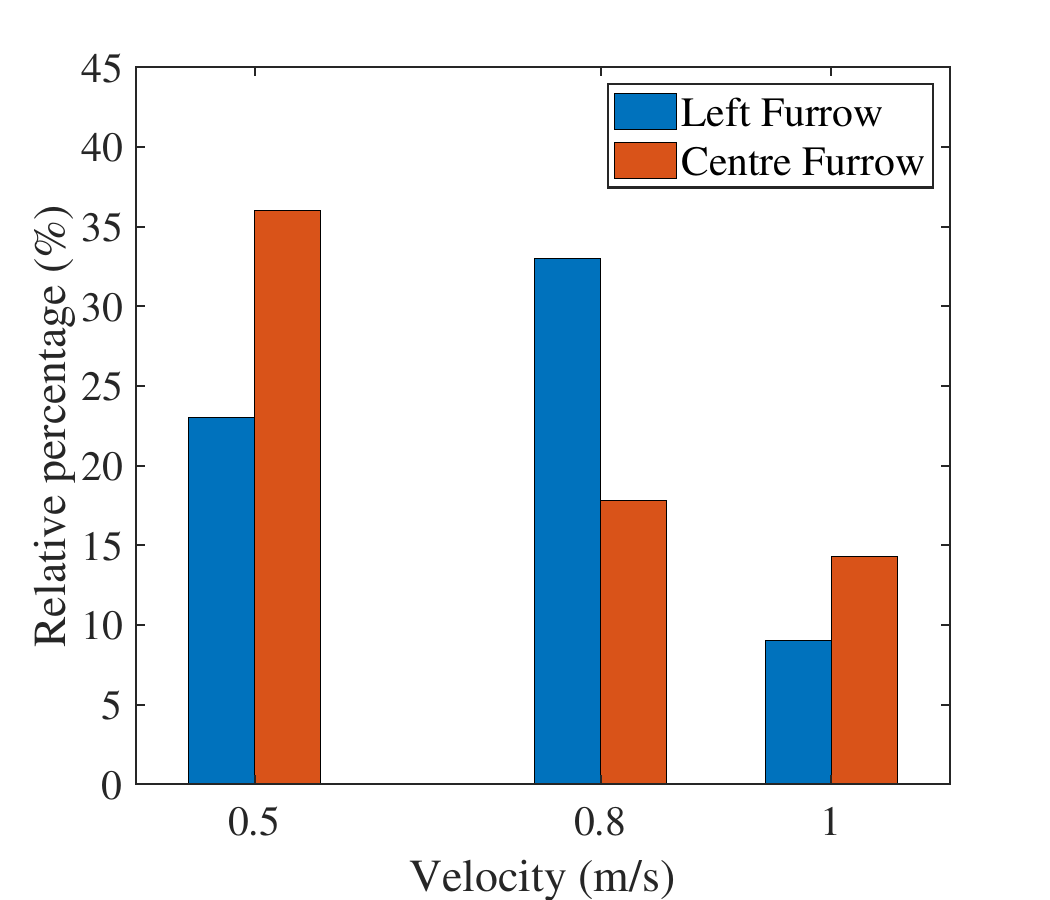}
\label{Fig:Relative increase in inversion index total}}
\caption{Relative increase in (a) horizontal, (b) vertical inversion index, and (c) combined inversion index}
\label{Fig:Relative increase in inversion index horizontal and vertical}
\end{figure}

The inversion index is determined by calculating the mass fraction transferred from one region to another. The inversion of soil layers depends on the vertical and horizontal movement of particles. Two furrows, i.e., left and centre, are studied for the left and centre plough, respectively. The initial and final distribution of soil particles is plotted using histograms along both the directions for the top and bottom layers, as shown in \Cref{Fig:Histograms_top_bottom_layer}. The mass fractions are determined for each region corresponding to the two furrows formed by the left and centre ploughs (see \Cref{tab:Mass_fractions}). Inversion efficiency has two components: horizontal and vertical. Therefore, relative changes in both components are studied the relative changes in both components are studied to study the effect of velocity on soil displacement in both directions.\\
For the left furrow, \SI{0.8}{\meter\per\second} of velocity shows the highest increase of 30.61\% and 35.9\% in both horizontal and vertical directions, respectively (see \Cref{Fig:Relative increase in inversion index horizontal,Fig:Relative increase in inversion index vertical}). Whereas, it is observed that for the centre furrow, 0.5 m/s velocity has shown the highest increment of 43.9\% and 27.5\% in inversion index among the other velocities. The modified design of the mouldboard plough has shown a significant increase in the inversion index for both the furrows at velocities of 0.5, 0.8 and \SI{1.0}{\meter\per\second}. From \Cref{Fig:Relative increase in inversion index total}, it can be observed that for the left furrow, the modified design has shown an increase of 23.08\%, 32.95\%, and 8.96\% in inversion index at velocities of 0.5, 0.8, and 1.0 m/s respectively, whereas for centre furrow it is 35.08\%, 17.81\%, and 14.29\%.
\begin{table}[htbp]
    \centering
    \caption{Mass fractions of displaced soil}
    \begin{tabular}{cclccccc}
    \hline
    \hline
    Vel (\SI{}{\meter\per\second})     & MB Plough                   & \multicolumn{1}{c}{Furrow} & $H_{T}$   & $H_{B}$   & $V_{T}$   & $V_{B}$   & Inversion   Index \\
    \hline
    \hline
    \multirow{4}{*}{0.5} & \multirow{2}{*}{Lit} & Furrow 1 (Left)            & 0.09 & 0.29 & 0.19 & 0.09 & 0.65              \\
     &                           & Furrow 2 (Centre) & 0.09 & 0.32 & 0.22 & 0.18 & 0.81 \\ 
     \cmidrule(lr){2-8}
     & \multirow{2}{*}{Mod} & Furrow 1 (Left)   & 0.08 & 0.4  & 0.19 & 0.12 & 0.8  \\
     &                           & Furrow 2 (Centre) & 0.1  & 0.49 & 0.24 & 0.27 & 1.1  \\
      \cmidrule(lr){1-8}
    \multirow{4}{*}{0.8} & \multirow{2}{*}{Lit} & Furrow 1 (Left)            & 0.1  & 0.39 & 0.17 & 0.22 & 0.88              \\
     &                           & Furrow 2 (Centre) & 0.12 & 0.33 & 0.19 & 0.1  & 0.73 \\
     \cmidrule(lr){2-8}
    
     & \multirow{2}{*}{Mod} & Furrow 1 (Left)   & 0.08 & 0.56 & 0.22 & 0.31 & 1.17 \\
     &                           & Furrow 2 (Centre) & 0.07 & 0.45 & 0.2  & 0.14 & 0.86 \\
    \cmidrule(lr){1-8}
    
    \multirow{4}{*}{1}   & \multirow{2}{*}{Lit} & Furrow 1 (Left)            & 0.27 & 0.29 & 0.32 & 0.47 & 1.34              \\
     &                           & Furrow 2 (Centre) & 0.12 & 0.81 & 0.2  & 0.49 & 1.61 \\
    \cmidrule(lr){2-8}
    
     & \multirow{2}{*}{Mod} & Furrow 1 (Left)   & 0.3  & 0.32 & 0.35 & 0.49 & 1.46 \\
     &                           & Furrow 2 (Centre) & 0.14 & 0.86 & 0.29 & 0.54 & 1.84 \\
    \hline
    \hline
    \end{tabular}
    
\label{tab:Mass_fractions}
\end{table}
%
\subsection{Wear analysis}\label{Sec:Wear}

The Archard wear model was employed to analyse the wear on the plough surface. This model defines wear volume as a function of the normal load, sliding distance, material hardness, and a dimensionless wear coefficient, and it is defined as \citep{archard1953contact},
\begin{equation}
Q = W F_n {\delta}_t
\end{equation}
where $Q$ is the volume of material removed, ${\delta}_t$ is the tangential distance moved and wear constant $W = K/H $, K is a dimensionless constant, and H is a hardness measure of the softest surface.

\begin{figure}[H]
\centering
\includegraphics[width=0.45\textwidth]{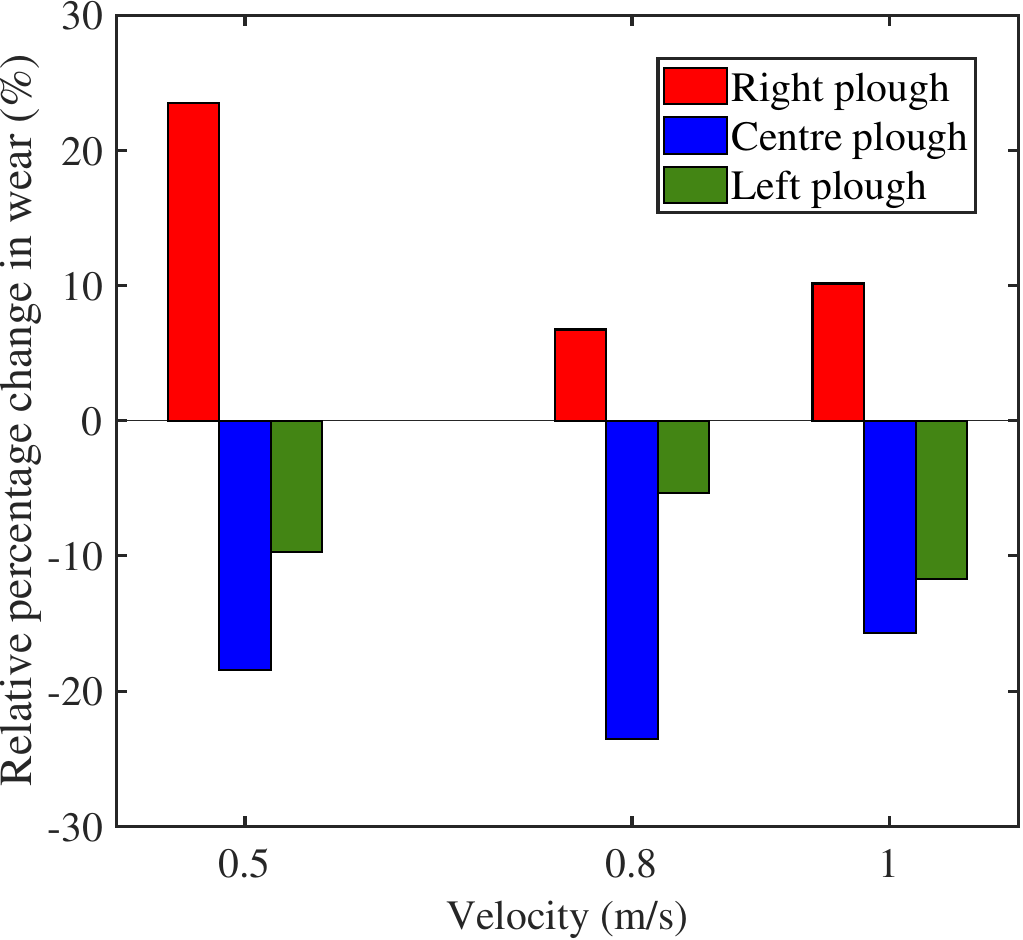}
\label{Fig: Relative percentage change in Wear for Modified Design}%
\caption{Relative percentage change in wear}
\label{Fig:P1 P2 P3 Wear and relative change}
\end{figure}

Archard wear is calculated for both designs, with the wear constant, $K$ is taken as \SI{e-8}{\pascal} approximately as mentioned by \citet{Napiorkowski2019}. The value of $K$ is used throughout the analysis to compare the relative wear rate between the MB ploughs. To compare the wear for both the designs, the relative percentage change over the literature design is calculated and shown in \Cref{Fig:P1 P2 P3 Wear and relative change}. The variation of wear with respect to distance for the tool travelling with a velocity of \SI{1}{\meter\per\second}. It is observed that the amount of material removed from the surface increases as the plough travels, and maximum wear is observed for the rightmost plough, whereas the least occurs for the left plough. This is due to the mass of soil coming in contact with the right plough being more than the left. Total wear after the ploughing for velocities of 0.5, 0.8 and \SI{1}{\meter\per\second} are given in \Cref{Tab:total wear}. There is an increase in wear by 23.49\%, 6.72\% and 10.13\% for velocities 0.5, 0.8 and \SI{1}{\meter\per\second}, respectively, for the right plough. The left and centre plough shows a significant decrease of 9\% to 23.70 \% in the amount of wear in modified design over literature. 
\begin{table}
    \centering
    \caption{Total wear on ploughs at different velocities}
    \begin{tabular}{cllll}
    \hline
    \hline
    \multirow{2}{*}{Velocity} & \multicolumn{1}{c}{\multirow{2}{*}{Tool}} & \multicolumn{3}{c}{Total Wear (mm)} \\
    \cmidrule(lr){3-5}
    & \multicolumn{1}{c}{} & Right plough   & Centre plough  & Left plough    \\
                             \hline
                             \hline
    \multirow{2}{*}{v = 0.5 m/s} & Literature           & 47.94 & 50.94  & 46.05 \\
                             & Modified             & 59.20 & 41.50 & 41.52 \\
    \multirow{2}{*}{v = 0.8 m/s} & Literature           & 48.38 & 50.43 & 34.72 \\
                             & Modified             & 51.63 & 38.48 & 32.82 \\
    \multirow{2}{*}{v = 1.0 m/s} & Literature           & 43.42 & 40.15 & 32.64  \\
                             & Modified             & 47.82 & 33.81 & 28.78 \\
    \hline
    \hline
    \end{tabular}
    
    \label{Tab:total wear}
\end{table}
%
\subsection{Stress distribution}\label{Sec:stress_distribution}
\begin{figure}[H]
         \centering
         \includegraphics[width=0.8\textwidth]{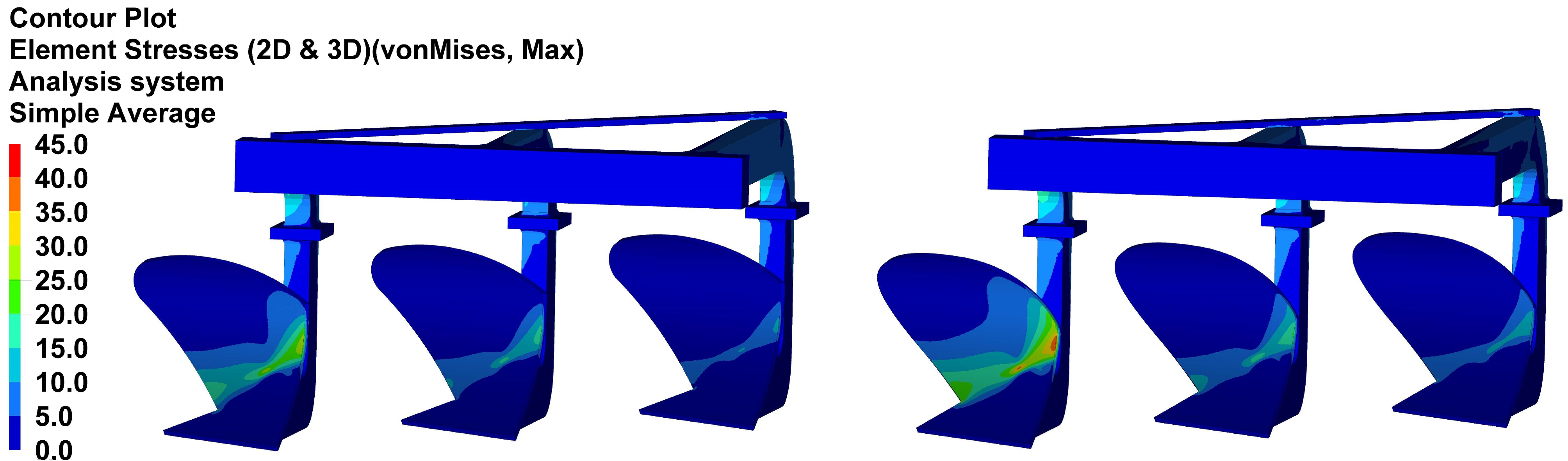}\put(-350,-15){(a) Literature MB plough}\put(-140,-15){(b) Modified MB plough}
         \caption{Stress distribution of plough at v = 0.5 m/s}
         \label{Fig:Stress_vel_0.5}
\end{figure}

\begin{figure}[H]
         \centering
         \includegraphics[width=0.95\textwidth]{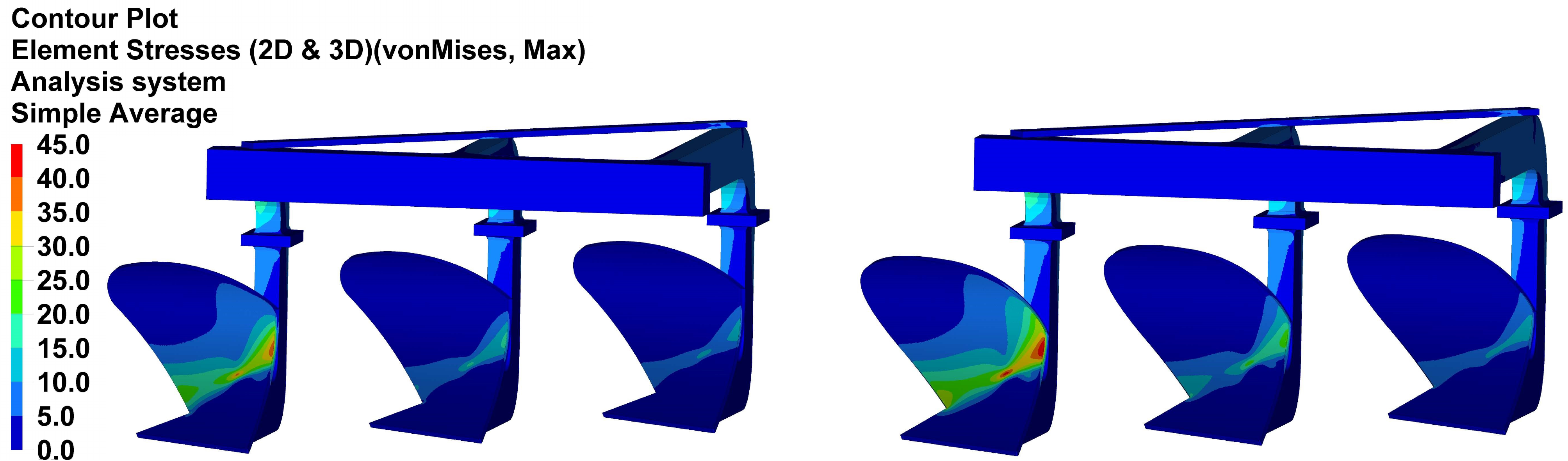}\put(-350,-15){(a) Literature MB plough}\put(-140,-15){(b) Modified MB plough}
         \caption{Stress Distribution of plough at v = 0.8 m/s}
         \label{Fig:Stress_vel_0.8}
\end{figure}

\begin{figure}[H]
         \centering
         \includegraphics[width=0.95\textwidth]{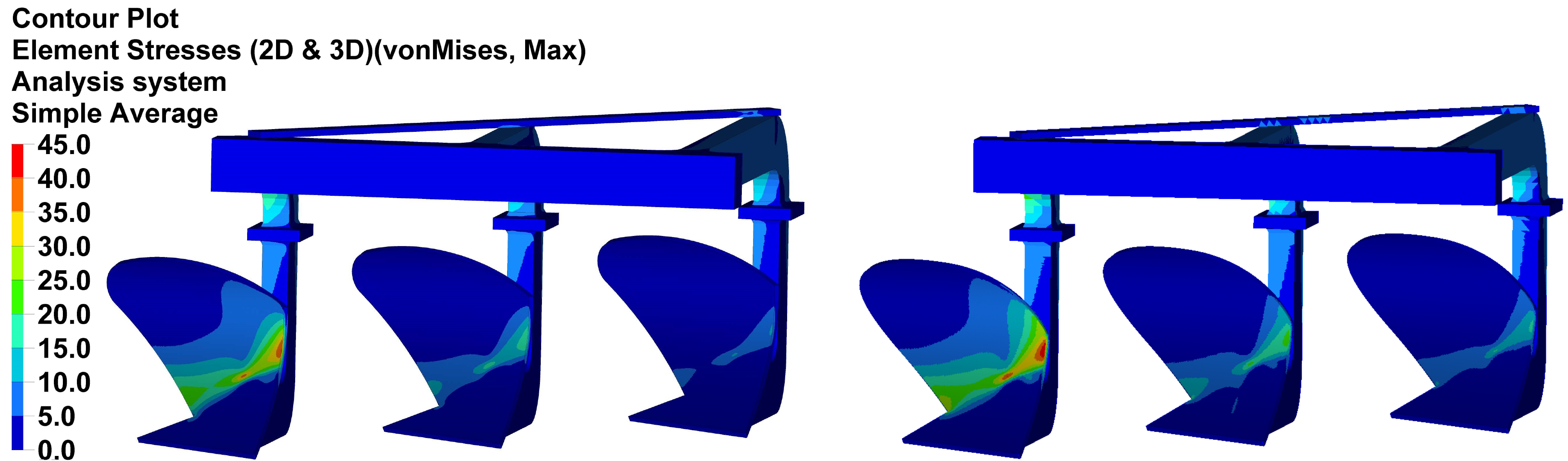}\put(-350,-15){(a) Literature MB plough}\put(-140,-15){(b) Modified MB plough}
         \caption{Stress Distribution of plough at v = 1.0 m/s}
         \label{Fig:Stress_vel_1.0}
\end{figure}

This section presents the stress variation profile on the surface of the MB plough due to the change in profile and velocity. Stresses are generated due to the contact forces exerted by the bulk of soil particles on the MB surface. Contact forces are extracted from DEM simulation in EDEM and mapped to the mesh in HyperMesh after meshing the geometry. The FEA analysis to estimate stresses is performed in OptiStruct \citep{Optistruct} and HyperView \citep{Hyperworks}. The plough geometry is meshed using tetrahedral second-order elements with a mesh size of \SI{0.5}{\milli\meter}. \Cref{Fig:Stress_vel_0.5,Fig:Stress_vel_0.8,Fig:Stress_vel_1.0} depict the Von Mises stress distribution on the surface of the three ploughs with velocities 0.5, 0.8 and \SI{1.0}{\meter\per\second}. As discussed in \Cref{sec:forces on MB plough}, the right plough experiences higher forces due to the increased lifting of soil, resulting in higher stresses in the right plough. The maximum stress induced is in the range of 30-47 \SI{}{\mega\pascal}, 16-25 \SI{}{\mega\pascal}, and 9-17 \SI{}{\mega\pascal} for right, centre, and left plough respectively which is much lower than the yield stress (\SI{240}{\mega\pascal}) of structural steel (see \Cref{tab: max stress induced}). Although the stresses induced in the modified design are higher than those in the literature design, they are still within safe limits, as they are well below the yield strength of the material.
\begin{table}
    \centering
    \caption{Maximum stress induced on each plough}
    \begin{tabular}{cllll}
    \hline
    \hline
        \multirow{2}{*}{\textbf{Velocity}} & \multirow{2}{*}{\textbf{Tool}} & \multicolumn{3}{c}{\textbf{Max Stress (MPa)}}  \\ 
        \cmidrule(lr){3-5}
         &  & Right & Centre & Left  \\ \hline \hline
        \multirow{2}{*}{0.5} & Literature & 30.15 & 18.19 & 11.62 \\ \cmidrule(lr){2-5}
        & Modified & 40.95 & 20.66 & 14.93  \\ \hline
        \multirow{2}{*}{0.8} & Literature & 39.43 & 17.36 & 13.26   \\ \cmidrule(lr){2-5}
         & Modified & 46.85 & 24.68 & 15.14  \\ \hline
        \multirow{2}{*}{1.0} & Literature & 40.37 & 19.36 & 10.87  \\ \cmidrule(lr){2-5}
         & Modified & 45.04 & 23.50 & 16.56 \\ 
    \hline
    \hline
    \end{tabular}
    
    \label{tab: max stress induced}
\end{table}
\section{Conclusions} \label{sec: Conclusions}
Ploughing simulations using a primary tillage tool were carried out using the DEM numerical technique in EDEM commercial software. The modification in the plough profile is made using a graphical method and compared to the earlier proposed design. The inversion efficiency, wear, forces and stresses generated in both designs were analysed. From the study, the following conclusions can be drawn.
\begin{enumerate}
    \item In both the designs, among the right, centre and left ploughs, the right plough experiences more draft force and contributes more to the power requirement. The force exerted by the soil particles on the ploughs is increased in the modified design by 15.28\%, 10.68\% and 7\% for the right, centre and left ploughs, respectively. This is because the modified plough has 6.6 \% more surface area than the literature one. Additionally, it was observed that draft force increases with the increase in velocity.
    
    \item A new method to calculate the performance of the tool to invert the soil has been proposed in this work, given the inversion index by calculating the horizontal and vertical inversion index, which allows us to study the movement of soil layers in both directions.

    \item Modified design shows better inversion efficiency than the literature design. At velocities of 0.5, 0.8 and \SI{1.0}{\meter\per\second}, the inversion index shows an increase of 23.08\%, 32.95\% and 9.96\% respectively, for left furrow and 35.08\%, 17.81\% and 14.29\% respectively, for the centre furrow. The right furrow is excluded from the analysis of the inversion index because the right plough is used to make the space for the soil to be turned by the centre plough. 
    
    \item Even though the surface area and mass of soil lifted is greater in the modified design, a significant decrease in wear of 9-23.70\% is observed in the left and right plough compared to the literature design. 
    
    \item A slight increase in stress is observed in the modified design as the forces acting on the surface are more. For both designs, the stresses induced are in the range of 30-47 \SI{}{\mega\pascal}, 16-25 \SI{}{\mega\pascal}, and 9-17 \SI{}{\mega\pascal} for the right, centre and left plough, respectively. The stresses induced are within the permissible limit as the yield stress for structural steel is 240 \SI{}{\mega\pascal}.
    
\end{enumerate}

While this study provides valuable insights into soil–tool interaction through simulation-based analysis, it has certain limitations. A reduced ploughing length was adopted to minimise computational cost; however, the tool and soil properties were kept at full scale to preserve the accuracy of force, wear, and stress predictions. Additionally, a simplified theoretical wear model was used, which does not account for complex field conditions such as moisture variation, heterogeneous soil composition, or dynamic tool wear mechanisms.

This study is based exclusively on simulation data. While field experiments or laboratory-scale physical validations have not been conducted at this stage, the findings offer valuable insights into soil–tool interactions through numerical modelling. The results are intended to guide the relative evaluation of design modifications and process parameters. Experimental validation under both controlled laboratory and field conditions is planned as part of future work to further substantiate and extend the applicability of these conclusions. Additionally in this study, a simplified theoretical wear model was employed to gain preliminary insights into wear trends under controlled conditions. However, real-world wear behavior is governed by a complex interplay of multiple factors, which will be addressed in future work.
\section*{Acknowledgements}
The authors thankfully acknowledge the generous research funding and simulation resources provided by Altair India Private Limited. The authors also acknowledge the generous financial support from IIT Madras under the Institutes of Eminence (IoE) scheme funded by the Ministry of Education, Government of India.
\section*{Conflict of Interest}
The authors declare that they have no conflict of interest.

\end{document}